\documentclass{llncs}
\pdfoutput=1
\usepackage{amsmath}
\usepackage{amssymb}
\usepackage{array}
\usepackage{hyperref}
\usepackage{listings}
\usepackage[nooneline,tight]{subfigure}
\usepackage{tikz}
\usepackage{booktabs}
\usepackage{microtype}
\usepackage{multirow}
\usepackage{flushend}
\usepackage{stmaryrd}
\usepackage[T1]{fontenc}
\usepackage{xspace}
\usepackage{color}
\usepackage{langs}
\usepackage{xparse}
\usepackage{subdepth}
\usepackage{wrapfig}

\usetikzlibrary{positioning,chains,shapes.arrows,shapes.geometric,fit,calc,arrows,decorations.pathmorphing}

\newcolumntype{C}[1]{>{\centering\arraybackslash}m{#1}} 
\newcolumntype{P}[1]{>{\arraybackslash}m{#1}} 

\newenvironment{itemize*}{%
  \begin{itemize}\addtolength{\itemsep}{-.35\baselineskip}}{%
  \end{itemize}}
\newenvironment{enumerate*}{%
  \begin{enumerate}\addtolength{\itemsep}{-.35\baselineskip}}{%
  \end{enumerate}}
\newenvironment{description*}{%
  \begin{description}\addtolength{\itemsep}{-.35\baselineskip}}{%
  \end{description}}

\makeatletter
\newcommand*{\mycleardoublepage}{\clearpage\if@twoside
\ifodd\c@page
  \hbox{}
  \clearpage
\fi
\hbox{}
\thispagestyle{empty}
\clearpage
\fi}
\makeatother

\newcounter{MOPReqCounter}

\makeatletter
\def\list@empty#1{}
\def\list@add#1#2{\edef#1##1{#1{##1}##1{#2}}}
\def\list@iterate#1#2{#1{#2}}
\makeatother

\makeatletter

\let\sf@oldsubfigure=\subfigure
\let\subfigure=\thisisnotdefined
\let\sfenv@beginSubfigure=\relax
\let\sfenv@endSubfigure=\relax
\let\sfenv@beforeSubfigure=\relax
\let\sfenv@afterSubfigure=\relax
\newbox\subfigbox 
\newenvironment{subfigure}
{\def\caption##1{\gdef\subcapsave{\relax##1}}%
\let\subcapsave=\@empty 
\let\sf@oldlabel=\label
\def\label##1{\global\list@add{\sublabsave}{##1}}
\let\sublabsave\list@empty 
\setbox\subfigbox\hbox\bgroup\sfenv@beginSubfigure}
{\sfenv@endSubfigure\egroup 
\let\label=\sf@oldlabel
\sfenv@beforeSubfigure%
\sf@oldsubfigure[\subcapsave\list@iterate\sublabsave\label]{\box\subfigbox}}%
\sfenv@afterSubfigure%
\makeatother

\makeatletter
{\hfil}
\makeatother

\makeatletter
{\hfil}
\makeatother

\makeatletter
{}
\makeatother

\newenvironment{indented}[1][]%
{\noindent
 \begin{itemize}
 \item[#1]}%
{\end{itemize}}

\newcounter{goal}{}
\definecolor{keyword}{RGB}{96,0,53}
\definecolor{darkgreen}{RGB}{0,128,0}
\definecolor{listingsFrame}{gray}{0.6}

\newcommand{\fboxgray}[1]{%
    \colorlet{currentcolor}{.}%
    {\color{listingsFrame}%
    \fbox{\color{currentcolor}#1}}%
}

\lstdefinestyle{normal}{%
  xleftmargin=\parindent,
}

\lstdefinestyle{nospace}{%
  aboveskip=0em,
  belowskip=0em,
  xleftmargin=0em,
  xrightmargin=0em,
}

\lstdefinestyle{nohspace}{%
  xleftmargin=0em,
  xrightmargin=0em,
}

\lstdefinestyle{scriptsize}{%
  basicstyle=\scriptsize\sf,
}

\lstdefinestyle{footnote}{%
  basicstyle=\footnotesize\sf,
}

\lstdefinestyle{figureframe}{%
  xleftmargin=0pt,
  frame=single,
  rulecolor=\color{listingsFrame},
  numbersep=6pt,
}
\lstdefinestyle{inlineframe}{%
  xleftmargin=2pt,
  xrightmargin=2pt,
  frame=single,
  rulecolor=\color{listingsFrame},
  numbersep=6pt,
}

\newcommand{\mytilde}{%
  \texttt{\resizebox{.48em}{1ex}{\hbox{$\sim$}}}%
}

\lstdefinelanguage{LaTeX}[]{Tex}{%
  language=Tex,
  morekeywords={emph, label, ref, begin, end, part, chapter, section,
                subsection, subsubsection, paragraph, subparagraph, cite},
}

\lstdefinelanguage{SugarJ}[]{Java}{%
  language=Java,
  morekeywords={sugar, extension, context, free, syntax, desugarings, sorts, signature,
    constructors, rules, strategies, assert, as, editor, services, colorer, folding,
    outliner, checks, recursive, errors, warnings, where,
    css, outlining, rec, color, completions, completion, template,
    layout, analyses, then},
  mathescape=true,
  deletestring=[b]',
  morecomment=[l]{//},
}

\lstdefinelanguage{SugarJXML}[]{SugarJ}{%
  deletecomment=[l]{//},
  morecomment=[s]{<!--}{-->}
}

\lstdefinelanguage{SugarMDD}[]{SugarJ}{%
  morekeywords={model, transformation},
}

\lstdefinelanguage{SugarJATM}[]{SugarMDD}{%
  morekeywords={statemachine, initial, state, events},
}

\lstdefinelanguage{SugarJEntity}[]{SugarMDD}{%
  morekeywords={entity},
}
\lstdefinelanguage{SugarJATMEntity}[]{SugarJATM}{%
  morekeywords={data, entity},
}

\lstdefinelanguage{SugarJTemplate}[]{SugarMDD}{%
  morekeywords={in, \$for, template},
  mathescape=false
}

\lstdefinelanguage{SugarJFeature}[]{SugarMDD}{%
  morekeywords={featuremodel, features, constraint, config, enable, disable, variable},
  otherkeywords={\#ifdef},
}

\lstdefinelanguage{scala}{
  morekeywords={abstract,case,catch,class,def,%
    do,else,extends,false,final,finally,%
    for,if,implicit,import,match,mixin,%
    new,null,object,override,package,%
    private,protected,requires,return,sealed,%
    super,this,throw,trait,true,try,%
    type,val,var,while,with,yield},
  sensitive=true,
  morecomment=[l]{//},
  morecomment=[n]{/*}{*/},
  morestring=[b]",
  morestring=[b]',
  morestring=[b]"""
}

\lstdefinelanguage{SDF}{%
  morekeywords={context, free, syntax, sorts, signature,
    constructors, rules, strategies, module, imports, exports},
  escapechar=_,
  mathescape=true,
  comment=[l]{\%\%},
  morestring=[b]",
}

\lstdefinelanguage{MyPython}[]{Python}{%
  escapechar=\%,
  mathescape=true,
}

\lstdefinelanguage{MyHaskell}[]{Haskell}{%
  escapechar=\%,
  mathescape=true,
  deletekeywords={Nothing,Just,False,True,putStrLn,fail,fromJust,lookup,Num,exp,free,snd,String,
  return,error,otherwise,not,show,read,Eval,Read,readsPrec,print},
}

\lstdefinelanguage{SugarHaskell}[]{MyHaskell}{%
  morekeywords={context, free, syntax, desugarings, sorts, signature,
    constructors, rules, strategies, lexical, reject},
  mathescape=false,
}

\lstdefinelanguage{SugarHaskellArrows}[]{SugarHaskell}{%
  morekeywords={proc},
}

\lstdefinelanguage{EBNF}{
  morestring=[b]"
}

\lstdefinelanguage{Constraint}{
}

\lstdefinelanguage{Plain}{}

\lstdefinelanguage{Pseudo}{
  keywords={foreach,match,case,in,return,if,else},
  mathescape=true,
}

\lstdefinelanguage{Questionnaire}[]{Java}{%
  morekeywords = {questionnaire, question, value, Boolean, String, Integer, group,
    if, else, define, ask}
}

\lstdefinelanguage{SugarFomega}{
  keywords = {module, val, type, mu, if, then, else, case, of,
    fold, unfold, true, false, as, public, import, syntax, desugaring, typing, context, free, let, in, end, forall, do},
  mathescape = true,
  morestring=[b]",
}

\usetikzlibrary{positioning,chains,fit,calc,arrows,decorations.pathmorphing}
\usetikzlibrary{shapes.arrows,shapes.geometric,shapes.symbols}

\tikzstyle{invisible} = []

\tikzstyle{model}
  = [ shape=rectangle
    , draw
    ]
\tikzstyle{transformation}
  = [ model
    , shape=single arrow
    , draw
    ]

\tikzstyle{meta} = [ double ]
\tikzstyle{generated} = [ dashed ]

\tikzstyle{mopdependency}
  = [ -stealth'
    , semithick
    ]
\tikzstyle{instance}
  = [ mopdependency
    , -open triangle 60
    ]

\tikzstyle{document}
  = [ shape=rectangle
    , draw
    , minimum height=1.5em
    , text width=5em
    , text centered
    ]

\tikzstyle{component}
  = [ font=\scriptsize\it
    ]

\tikzstyle{code}
  = [ shape=rectangle
    , draw
    ]

\tikzstyle{process}
  = [ shape=single arrow
    , single arrow head extend=.75em
    , single arrow head indent=.25em
    , minimum width=3em
    , draw
    ]

\tikzstyle{point}
  = [ coordinate
    , minimum width=1em
    ]

\tikzstyle{flow diagram}
  = [ start chain
    , node distance=1em
    , every node/.style={on chain}
    ]

\tikzstyle{ast}
  = [ node distance=1em and 0.38em
    , every node/.style=ast node
    ]

\tikzstyle{ast node}
  = [ shape=circle
    , minimum size=0.5em
    , inner sep=0
    , fill
    ]

\tikzstyle{dependency}
  = [ dashed
    , -stealth'
    ]

\tikzstyle{red}
  = [ shape=rectangle
    , color=red
    ]

\tikzstyle{blue}
  = [ shape=circle
    , color=blue
    ]

\tikzstyle{green}
  = [ shape=diamond
    , color=green!80!black
    , minimum size=0.7em
    ]

\tikzstyle{note}
  = [ font=\scriptsize\it
    ]
    
\tikzstyle{zoomed arrow}
  = [ solid
    , decorate
    , decoration=snake
    , -stealth'
    ]

\tikzstyle{zoom}
  = [ dashed
    ]

\tikzstyle{uml class}
  = [ shape=rectangle
    , draw
    , font=\sf
    ]

\tikzstyle{uml package}
  = [ uml class
    , inner sep=1em
    ]

\tikzstyle{uml dependency}
  = [ dependency, 
    , dashed
    ]

\tikzstyle{double arrow}
  = [ shape=double arrow
    , double arrow head extend=.75em
    , double arrow head indent=.25em
    , minimum width=3em
    , draw
    , font=\sf
    ]

\newdimen\nodedistance
\tikzstyle{name graph}
  = [ node distance=\nodedistance
    , every node/.style={namenode}
    , every path/.style={ref}
    ]

\tikzstyle{namenode}
  = [ circle
    , thick
    , anchor=center
    , draw
    , minimum size=2em
    , inner sep=2pt
    ]

\tikzstyle{synthesized}
  = [ namenode
    , fill=gray!45
    ]

\tikzstyle{ref}
  = [ -stealth'
    , semithick
    , draw
    ]

\tikzstyle{badref}
  = [ ref
    , dashed
    ]

\newcommand{\distanceTop}{7.15pt}

\newcommand{\distanceBottom}{-2.55pt}

\newcommand{\distanceLeft}{-0.5pt}

\newcommand{\distanceRight}{0.5pt}

\newcommand{\drawrect}[4][]{\begin{tikzpicture}[remember picture, overlay]
\draw[layout box, #1]
  ($(#2) +(\distanceLeft, \distanceTop) + (-#4, #4)$) rectangle
  ($(#3) + (\distanceRight, \distanceBottom) + (#4, -#4)$);
\end{tikzpicture}}

\newcommand{\minirect}{\hbox to 9pt{\drawrect[thin,scale=0.3,black]{0pt, 15pt}{26pt, -5pt}{0pt}}}

\tikzstyle{layout box}
  = [ blue
    , semithick
    , draw
    ]

\tikzstyle{mini layout box}
  = [ black
    , very thin
    , draw
    ]

\tikzstyle{annotation}
  = [ font=\small\it
    ]

\tikzstyle{code annotation}
  = [ font=\small\it
    ]

\newcommand{\selectKeywordFont}{\bfseries}

\newcommand{\selectStringLitFont}{\tt}

\newcommand{\selectIdentifierFont}{\relax}

\newcommand{\selectOperatorFont}{\tt}

\newcommand{\tokenHeight}{2ex}

\newcommand{\tokenDepth}{.5ex}

\tikzstyle{pretty print}
  = [ start chain=going base right
    , text height=\tokenHeight
    , text depth=\tokenDepth
    , inner sep=0
    , node distance=0em
    ]

\tikzstyle{new line/helper}
  = [ on chain
    , anchor=north west,
      at={($(#1.west |- \tikzchainprevious.base) + (0, -1.1\baselineskip)$)}
    ]

\tikzstyle{new line tight/helper}
  = [ on chain
    , anchor=north west,
      at={($(#1.west |- \tikzchainprevious.base) + (0, -0.9\baselineskip)$)}
    ]

\tikzstyle{new line}[\tikzchainprevious]
  = [ on chain=placed {new line/helper=#1}
    ]

\tikzstyle{new line tight}[\tikzchainprevious]
  = [ on chain=placed {new line tight/helper=#1}
    ]

\tikzstyle{tab forward/helper}
  = [ on chain
    , anchor=north west,
      at={(#1 |- \tikzchainprevious.base)}
    ]

\tikzstyle{tab forward}
  = [ on chain=placed {tab forward/helper=#1}
    ]

\tikzstyle{indent}[1em]
  = [ on chain=placed {base right=#1 of \tikzchainprevious}
    ]

\tikzstyle{identifier}
  = [ on chain
    , font=\selectIdentifierFont
    ]

\tikzstyle{operator}
  = [ on chain
    , font=\selectOperatorFont
    ]

\tikzstyle{keyword}
  = [ on chain
    , font=\selectKeywordFont
    , text=keyword
    ]

\tikzstyle{stringlit}
  = [ on chain
    , font=\selectStringLitFont
    , text=blue
    ]

\newcommand{\Empty}{}
\newcommand{\ignore}[1]{\relax}

\makeatletter
\newcommand{\ppBox}[2][]{{
  \newcommand{\KW}[1]{
    \node[keyword] {##1};
    \HandleToken{\tikzchaincurrent}}

  \newcommand{\ID}[1]{
    \node[identifier] {##1};
    \HandleToken{\tikzchaincurrent}}

  \newcommand{\OP}[1]{
    \node[operator] {##1};
    \HandleToken{\tikzchaincurrent}}

  \newcommand{\SP}{
    \node[on chain] { };
    \HandleInsensitiveToken{\tikzchaincurrent}}

  \newcommand{\LB}[1]{
    \coordinate[on chain] (##1);}

  \newcommand{\TB}[1]{
    \coordinate[tab forward=##1];
  }

  \newcommand{\STR}[1]{
    \node[stringlit] {##1};
    \HandleToken{\tikzchaincurrent}}

  \renewcommand{\\}{
    \coordinate[new line=\NewlineNode];
    \let\HandleToken=\HandleTokenSubsequentLine}

  \newcommand{\newlineTight}{
    \coordinate[new line tight=\NewlineNode];
    \let\HandleToken=\HandleTokenSubsequentLine}

  \let\FirstToken=\Empty
  \newcommand{\AdjustFirst}[1]{
    \ifx\FirstToken\Empty
    \edef\FirstToken{##1}
    \fi
  }

  \let\LastToken=\Empty
  \newcommand{\AdjustLast}[1]{
    \edef\LastToken{##1}
  }

  \let\LeftToken=\Empty
  \newcommand{\AdjustLeft}[1]{
    \ifx\LeftToken\Empty
      \edef\LeftToken{##1}
    \else
      \pgf@process{\pgfpointanchor{##1}{west}}
      \setlength{\pgf@xa}{\pgf@x}
      \pgf@process{\pgfpointanchor{\LeftToken}{west}}
      \setlength{\pgf@xb}{\pgf@x}
      \ifdim\pgf@xa<\pgf@xb
        \edef\LeftToken{##1}
      \fi
    \fi
  }

  \let\RightToken=\Empty
  \newcommand{\AdjustRight}[1]{
    \ifx\RightToken\Empty
      \edef\RightToken{##1}
    \else
      \pgf@process{\pgfpointanchor{##1}{east}}
      \setlength{\pgf@xa}{\pgf@x}
      \pgf@process{\pgfpointanchor{\RightToken}{east}}
      \setlength{\pgf@xb}{\pgf@x}
      \ifdim\pgf@xa>\pgf@xb
        \edef\RightToken{##1}
      \fi
    \fi
  }

  \newcommand{\HandleTokenFirstLine}[1]{
    \AdjustFirst{##1}
    \AdjustRight{##1}
    \AdjustLast{##1}
  }

  \newcommand{\HandleTokenSubsequentLine}[1]{
    \AdjustLeft{##1}
    \AdjustRight{##1}
    \AdjustLast{##1}}

  \newcommand{\HandleInsensitiveToken}[1]{
  }

  \newcommand{\HandleToken}[1]{}

  \newcommand{\ppSubBox}[2][]{
    \coordinate[on chain];
    {
      \let\NewlineNode\tikzchaincurrent

      \let\FirstToken=\Empty
      \let\LastToken=\Empty
      \let\LeftToken=\Empty
      \let\RightToken=\Empty

      \let\HandleToken=\HandleTokenFirstLine

      ##2

      \ifx\LeftToken\Empty
        \path [##1]
          (\FirstToken.north west) rectangle
          (\LastToken.south east);
      \else
        \path [##1]
          (\RightToken.east |- \FirstToken.north) --
          (\FirstToken.north west) --
          (\FirstToken.south west) --
          (\LeftToken.west |- \FirstToken.south) --
          (\LeftToken.west |- \LastToken.south) --
          (\LastToken.south east) --
          (\LastToken.north east) --
          (\RightToken.east |- \LastToken.north) --
          cycle;
      \fi

      \global\let\InnerFirstToken=\FirstToken
      \global\let\InnerLastToken=\LastToken
      \global\let\InnerLeftToken=\LeftToken
      \global\let\InnerRightToken=\RightToken

      \global\let\InnerHandleToken=\HandleToken
    }
    \HandleToken{\InnerFirstToken}
    \ifx\HandleToken\HandleTokenFirstLine
    \let\HandleToken=\InnerHandleToken
    \fi
    \ifx\InnerLeftToken\Empty
    \else
    \HandleToken{\InnerLeftToken}
    \fi
    \HandleToken{\InnerRightToken}
    \HandleToken{\InnerLastToken}
  }

  \newcommand{\IN}[1]{\coordinate[indent]; ##1}
  \newcommand{\DE}[1]{\coordinate[indent=-1em]; ##1}

  \let\ppBox=\ppSubBox
  \ppBox[#1]{#2}}}
\makeatother

\newif\iftechreport
\techreporttrue 

\newcommand{\captionskip}{1.5ex}
\def\irascal#1{\lstinline[language=rascal]{#1}}
\def\ijava#1{\lstinline[language=java]{#1}}

\begin{document}

\makeatletter \spn@wtheorem{convention}{Convention}{\bfseries}{\itshape} \makeatother
\makeatletter \spn@wtheorem{assumption}{Assumption}{\bfseries}{\itshape} \makeatother

\newcommand{\projecturl}{\url{http://sugarj.org}}

\newcommand{\resolve}[1]{\ensuremath{\mathit{resolve}^{\!#1}\!}\xspace}
\newcommand{\fixCapture}{\ensuremath{\mathit{name\textit{-}fix}}\xspace}
\newcommand{\rename}{\ensuremath{\mathit{rename}}\xspace}
\newcommand{\badBindings}{\ensuremath{\mathit{find\textit{-}capture}}\xspace}
\newcommand{\compRenamings}{\ensuremath{\mathit{comp\textit{-}renaming}}\xspace}

\newcommand{\nameat}[2]{\ensuremath{#2^{@#1}}}

\newcommand{\equivL}{\equiv_\labels}
\newcommand{\equivA}{\equiv_\alpha}
\newcommand{\equivSA}[1]{\equiv_\alpha^{#1}}

\newcommand{\labels}{\ensuremath{\mathcal{L}}\xspace}
\newcommand{\pto}{\to}
\newcommand{\vref}{v_r}
\newcommand{\vdef}{v_d}

\newcommand{\tstring}{\textnormal{\textsf{string}}} 
\newcommand{\dom}{\ensuremath{\mathit{dom}}\xspace}
\newcommand{\codom}{\ensuremath{\mathit{codom}}\xspace}
\newcommand{\pisrc}{\ensuremath{\pi_{\textit{src}}}}
\newcommand{\pisyn}{\ensuremath{\pi_{\textit{syn}}}}

\newcommand{\tick}{\text{'}}

\def\highlight#1{\underline{\color{blue}{#1}}}
\def\shighlight#1{\underline{\color{blue}\tt{#1}}}

\newcommand{\insnode}[1]{\tikz[overlay,remember picture] \node (#1) {};}

\renewcommand{\paragraph}[1]{\vspace{.25\baselineskip}\noindent\textit{#1}}

\title{Capture-Avoiding and Hygienic\\ Program Transformations \iftechreport {\large (incl. Proofs)} \fi}
%

\author{Sebastian Erdweg\inst{1} \and Tijs van der Storm\inst{2,3}
  \and Yi Dai\inst{4}}
\institute{$^1$ TU Darmstadt, Germany \quad 
           $^2$ CWI, Amsterdam, The Netherlands \\
           $^3$ INRIA Lille, France \quad
           $^4$ University of Marburg, Germany}

\maketitle

\begin{abstract}
  Program transformations in terms of abstract syntax trees compromise
  referential integrity by introducing variable capture. Variable capture occurs
  when in the generated program a variable declaration accidentally shadows the
  intended target of a variable reference. Existing transformation systems
  either do not guarantee the avoidance of variable capture or impair the
  implementation of transformations.

%
%
  We present an algorithm called \fixCapture that automatically eliminates
  variable capture from a generated program by systematically renaming
  variables.  \fixCapture is guided by a graph representation of the binding
  structure of a program, and requires name-resolution algorithms for the source
  language and the target language of a transformation.  \fixCapture is generic
  and works for arbitrary transformations in any transformation system that
  supports origin tracking for names.  We verify the correctness of \fixCapture
  and identify an interesting class of transformations for which \fixCapture
  provides hygiene. We demonstrate the applicability of \fixCapture for
  implementing capture-avoiding substitution, inlining, lambda lifting, and
  compilers for two domain-specific languages.
\end{abstract}

\section{Introduction}
\label{sec:introduction}

Program transformations find ubiquitous application in compiler construction to
realize desugarings, optimizers, and code generators. While traditionally the
implementation of compilers was reserved for a selected few experts, the current
trend of domain-specific and extensible programming languages exposes developers
to the challenges of writing program transformations. In this paper, we
address one of these challenges: capture avoidance.

A program transformation translates programs from a source language to a target
language. In doing so, many transformations reuse the names that occur in a
source program to identify the corresponding artifacts generated in the target
program. For example, consider the compilation of a state machine to a
simple procedural language as illustrated in
Figure~\ref{fig:door-state-machine-simple}. The state machine has three states
\lstinline!opened!, \lstinline!closed!, and \lstinline!locked!. For each state
the compiler generates a constant integer function with the same name. Furthermore, for each state the compiler generates a dispatch function
that takes an event and depending on the event returns the subsequent state. For example, the dispatch
function for \lstinline!opened! tests if the given event is \lstinline!close!
and either yields the integer constant representing the following state
\lstinline!closed! or a dynamic error. Finally, the compiler generates a
main dispatch function that calls the dispatch function of the current state.

\begin{figure}[tp]
  \begin{subfigure}
    \begin{minipage}{.25\linewidth}
\begin{lstlisting}[language=statemachine,style=figureframe]
state opened
˚˚close  =>  closed

state closed
˚˚lock  =>  locked
˚˚open  =>  opened

state locked
˚˚unlock  =>  closed



˚
\end{lstlisting}
    \end{minipage}
    \caption{Door state machine.}
  \label{fig:door-state-machine}
  \end{subfigure}
\hfill
  \begin{subfigure}
    \begin{minipage}{.65\linewidth}
\begin{lstlisting}[language=simple,style=figureframe,numbers=left]
fun opened() =  0;
fun closed() =  1;
fun locked() =  2;
fun opened-dispatch(event) =
˚˚if (event == "close") then closed() else error();
fun closed-dispatch(event) =
˚˚if (event == "open") then opened()
˚˚else if (event == "lock") then locked() else error();
fun locked-dispatch(event) =
˚˚if (event == "unlock") then closed() else error();
fun main-dispatch-next-event(state, event) =
˚˚if (state == opened()) then opened-dispatch(event)
˚˚else if (state == closed()) [...];
\end{lstlisting}
    \end{minipage}
    \caption{Program generated for the door state machine.}
  \label{fig:door-simple}
  \end{subfigure}
  \caption{Many transformations reuse names from the source program in generated code.}
  \label{fig:door-state-machine-simple}
\end{figure}

A naive implementation of such compiler is easy to implement, but also runs the
risk of introducing variable capture. For example, if we consistently rename the state
\lstinline!locked! to \lstinline!opened-dispatch! as shown in
Figure~\ref{fig:door-state-machine-renamed}, we expect the compiler to produce
code that behaves the same as the code generated for the state machine without
renaming. However, a naive compiler blindly copies the state names into the
generated program, which leads to the incorrect code shown in
Figure~\ref{fig:door-simple-renamed}: The function definition on line~$4$
shadows the constant function on line~$3$ and thus captures the variable
reference \lstinline!opened-dispatch! on line~$8$ (we assume there is no
overloading). For the example shown, the problem is easy
to fix by renaming the dispatch function on line~$4$ and its reference on
line~$12$ to a fresh name \lstinline!opened-dispatch-0!. However, a general
solution is difficult to obtain. Existing approaches either rely on naming
conventions and fail to guarantee capture avoidance, or they
require a specific transformation engine and
affect the implementation of transformations.

\begin{figure}[tp]
  \begin{subfigure}
    \begin{minipage}{.25\linewidth}
\begin{lstlisting}[language=statemachine,style=figureframe,escapeinside=\`\`]
state opened
˚˚close  =>  closed

state closed
lock=>`\highlight{opened-dispatch}`
˚˚open  =>  opened

state `\highlight{opened-dispatch}`
˚˚unlock  =>  closed



˚
\end{lstlisting}
    \end{minipage}
    \caption{Consistently renaming door state machine.}
\label{fig:door-state-machine-renamed}
  \end{subfigure}
\hfill
  \begin{subfigure}
    \begin{minipage}{.65\linewidth}
\begin{lstlisting}[language=simple,style=figureframe,numbers=left,escapeinside=\`\`]
fun opened() =  0;
fun closed() =  1;
fun `\highlight{opened-dispatch}`() =  2;
fun `\highlight{opened-dispatch}`(event) =
˚˚if (event == "close") then closed() else error();
fun closed-dispatch(event) =
˚˚if (event == "open") then opened()
˚˚else if (event == "lock") then `\highlight{opened\hbox{-}dispatch}`() else ...
fun opened-dispatch-dispatch(event) =
˚˚if (event == "unlock") then closed() else error();
fun main-dispatch-next-event(state, event) =
˚˚if (state == opened()) then `\highlight{opened-dispatch}`(event)
˚˚else if (state == closed()) [...];
\end{lstlisting}
    \end{minipage}
    \caption{Program generated for the renamed door state machine is incorrect:
      Variable capture of \lstinline!opened-dispatch!.}
  \label{fig:door-simple-renamed}
  \end{subfigure}
  \caption{Variable capture can occur when original and synthesized names
    are mixed.}
  \label{fig:door-state-machine-simple-renamed}
\end{figure}

We propose a generic solution called \fixCapture that guarantees capture
avoidance and does not affect the implementation of transformations. \fixCapture
compares the name graph of the source program with the name graph of the
generated program to identify variable capture. If there is variable capture,
\fixCapture systematically and globally renames variable names to differentiate the captured
variables from the capturing variables, while preserving intended variable
references among original variables and among synthesized variables,
respectively. \fixCapture requires name analyses for the source and target
languages, which often exists or are needed anyway (e.g., for editor services,
error checking, or refactoring), and hence can be reused. \fixCapture treats transformations as a black box and is independent of the
used transformation engine as long as it supports origin tracking for names~\cite{VanDeursenKT93}.

\fixCapture enables developers of program transformations to focus on the actual
translation logic and to ignore variable capture. In particular, \fixCapture
enables developers to use simple naming schemes for synthesized variables in the
transformation and to produce intermediate open terms. For example, in
Figure~\ref{fig:door-state-machine-simple}, we append
\lstinline!"-dispatch"$$! to a state's name to derive the name of the
corresponding dispatch function. This construction occurs at two independent
places in the transformation: When generating a dispatch function for a state,
and when generating the main dispatch function. The connection between these is
only established when assembling all parts of the generated program in the final
step of the transformation. Using \fixCapture, it is safe to apply global naming
schemes with intermediate open terms to associate generated variable references and
declarations. Transformations of this kind fall into the class of transformations
for which \fixCapture guarantees hygiene, that is, $\alpha$-equivalent source
programs are always mapped to $\alpha$-equivalent target programs.

In summary, we make the following contributions:

\begin{itemize}
\item We studied 9 existing DSL implementations that use transformations and found that
  8 of them were prone to variable capture.
\item We present \fixCapture, an algorithm that automatically eliminates variable capture from
  the result of a program transformation.
\item We state and verify termination and correctness properties for \fixCapture and
  show that \fixCapture produces $\alpha$-equivalent programs for programs
  that are equal up to consistent but possibly capturing renaming.
\item We propose a notion of hygienic transformations and identify an interesting class of
  transformations for which \fixCapture provides hygiene.
\item We present an implementation of \fixCapture in the metaprogramming system
  Rascal. Our implementation supports capture avoidance for transformations that
  generate code as syntax trees or as strings.
\item We demonstrate the applicability of \fixCapture in a wide range of scenarios:
  for capture-avoiding substitution, for optimization (function inlining), for desugaring of language extensions
  (lambda lifting), and for code generation (compilation of DSLs for state
  machines and for digital forensics).
\end{itemize}

\addtolength{\belowcaptionskip}{\captionskip}

\section{Capture-avoiding transformations: What and why}
\label{sec:state-machine-example}

Capture avoidance is best known from capture-avoiding substitution: When
substituting an expression $e_2$ under a binder as in $\lambda\,x.\
(e_1[y:=e_2])$, variable $x$ may not occur free in $e_2$ otherwise the original
binding of $x$ in $e_2$ would be shadowed by the $\lambda$. To implement
capture-avoiding substitution, we must rename $x$ to a fresh variable $\alpha
\not\in \{y\} \cup \textit{FV}(e_1) \cup \textit{FV}(e_2)$ to avoid the capture:
$\lambda\,\alpha.\ (e_1[x:=\alpha][y:=e_2])$. Ensuring capture avoidance is
already relatively complicated for substitution in the $\lambda$-calculus. For
larger languages and more complex program transformations, ensuring capture
avoidance is a non-trivial and error-prone task.

\subsection{Variable capture in the wild}

To better understand the relevance of the problem of variable capture, we
studied implementations of a DSL for questionnaires in 10
state-of-the-art language workbenches in the context of the Language Workbench
Challenge 2013~\cite{ErdwegSV13}.\footnote{We studied all workbenches
  of the previous study~\cite{ErdwegSV13}: Ens\=o,
  M\'as, MetaEdit+, MPS, Onion, Rascal, Spoofax, SugarJ, the Whole Platform, and
  Xtext.} The questionnaire DSL features named declarations of questions and
named definitions of derived values. 9 of the 10 language workbenches translate a
questionnaire into a graphical representation using either Java or HTML with CSS
and JavaScript as target language. One workbench uses interpretation instead of
transformation. In most cases, the implementation of the DSL was conducted by
the developers of the workbench themselves.

The result of our study is shocking: The DSL implementations in 8 of the 9
language workbenches that use transformations fail to address capture avoidance
and produce incorrect code even for minimal changes to the definition of a questionnaire. For
example, some implementations fail when a question name is changed to
\lstinline!container!, \lstinline!questions!, or \lstinline!SWTUtils!, because
these names are implicitly reserved for synthesized variables. Other
implementations of the DSL use naming schemes similar to the one we illustrated
in the state-machine example. If there is already a question called
\lstinline!Q!, these implementations fail when naming another question
\lstinline!QBlock!, \lstinline!calculated_Q!, or \lstinline!grp_Q!. Some of the
variable captures result in compile-time errors of the generated Java code,
others result in misbehaved code that, for example, silently skips some of the
questions when storing answers persistently. Debugging such errors typically
requires investigation of the generated code and can be very time-consuming.

Of the studied DSL implementations, only the transformation
implemented in M\'as addressed variable capture. It uses global name
mappings to generate unique names from source-language variables for
the generated code. The usage of these name mappings and similar approaches is cross-cutting
and relies on the discipline of the developer; it is not enforced or
supported by the framework. We seek a solution that provides stronger
guarantees and has less impact on the implementation of a
transformation.

\subsection{Problem statement}
\label{SECT:problemstatement}

The goal of this work is to provide a mechanism that avoids variable capture in
code that is generated by program transformations. To this end, we seek a
mechanism that satisfies the following design goals:

\begin{itemize}
\item[G1:] Preserve reference intent: If a reference from the
  source program occurs in the target program, then the original declaration must
  also occur in the target program and the reference is still bound by it. In
  other words, source-program variables may neither be captured by synthesized
  declarations nor by other source-program declarations.
\item[G2:] Preserve declaration extent: If a declaration from the
  source program occurs in the target program, then only source-program
  references may be bound by it. In other words, synthesized variable references
  may not be captured by source-program declarations.
\item[G3:] Noninvasive: Avoidance of variable capture should not impact the
  readability of generated code. This is important in practice, where the
  generated code is often manually inspected when debugging a program
  transformation. In particular, a generated program should be left unchanged if
  it does \emph{not} contain variable capture.
\item[G4:] Language-parametric: It should be possible to eliminate variable
  capture from virtually all source and target languages that feature static name resolution.
\item[G5:] Transformation-parametric: The mechanism should work with different
  transformation engines and should not impose a specific style of
  transforming programs. Ideally, the mechanism supports
  existing transformations unchanged.
\end{itemize}

\noindent In the following sections, we present our solution
\fixCapture. It fully achieves the first three goals. In addition,
\fixCapture is language-parametric provided the name
analysis of source and target language satisfy modest assumptions. Finally, \fixCapture
works with any transformation engine that provides origin tracking~\cite{VanDeursenKT93} for
variable names, so that names originating from the source program can be
distinguished from names synthesized by the transformation.

\section{Graph-guided elimination of variable capture}
\label{sec:namefix-explanation}

The core idea of our solution is to provide a generic mechanism for
the detection and elimination of variable capture based on name graphs
of the source and target program. We use the term \emph{name} for the
string-valued entity that occurs in the abstract syntax tree of a program.
Naturally, the same name may occur at multiple locations of a program.
To distinguish different occurrences of the same name, we assume names
are labeled with a variable ID. In source programs, such IDs are unique.
However, for target programs generated by some transformation, we do
not require that variable IDs are unique, because the transformation may
have copied and duplicated names from the input program to the output
program.

We write $x^v$ to denote that name $x$ is labeled with variable ID $v$,
and we write $\nameat{v}{p}$ to retrieve from program $p$ the name
corresponding to variable ID~$v$. Nodes that share the same ID must have
the same name so that $\nameat{v}{p}$ is uniquely determined. The
nodes of a name graph are the variable IDs that occur in a program and the
edges connect references to the corresponding declarations.

\vspace{-1ex}
\begin{definition}\label{def:name-graph}
\textnormal{The \emph{name graph} of a program $p$ is a pair $G = (V,\rho)$ where}
\par\noindent\hskip1em
$
\begin{array}{cl@{\hskip1em}l}
  V & & \textnormal{is the set of variable IDs in $p$ (references and declarations),} \\
  \rho & \in V \pto V & \textnormal{is a partial function from references to declarations,} \\
\end{array}
$
\par
\noindent \textnormal{and if $\rho(\vref) = \vdef$, then reference and
  declaration have the same name $\nameat{\vref}{p} \!=
  \nameat{\vdef}{p}$. }
\end{definition}
\vspace{-1ex}

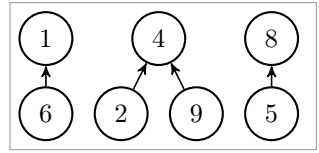
\begin{wrapfigure}{R}{.33\linewidth}
\vspace{-5ex}
\fboxgray{
    \begin{tikzpicture}[name graph]
      \node (dec-opened) at (1,2) {1}; 
      \node (dec-closed) at (2.5,2) {4};
      \node (dec-locked) at (4,2) {8};

      \node (ref-opened) at (1,1) {6};
      \node (ref-closed1) at (2,1) {2};
      \node (ref-closed2) at (3,1) {9};
      \node (ref-locked) at (4,1) {5};

     \path (ref-opened) edge (dec-opened);
     \path (ref-closed1) edge (dec-closed);
     \path (ref-closed2) edge (dec-closed);
     \path (ref-locked) edge (dec-locked);
    \end{tikzpicture}
}
\caption{Name graph of state machine in Figure~\ref{fig:door-state-machine}.}
  \label{fig:door-state-machine-graph}
\vspace{-5ex}
\end{wrapfigure}

\noindent For example, Figure~\ref{fig:door-state-machine-graph}
displays the name graph of the state machine in
Figure~\ref{fig:door-state-machine}, where we use line numbers as
variable IDs: ID 1 represents the declaration of \lstinline!opened!,
ID 2 represents the reference to \lstinline!closed! in the transition
on line 2, ID 4 represents the declaration of \lstinline!closed!, and
so on. 




We require that transformations preserve variable IDs when reusing names
from the source program in the generated code. For example, when compiling the state machine of
Figure~\ref{fig:door-state-machine} to the code in Figure~\ref{fig:door-simple},
the compiler reuses the names of state declarations for the declaration of
constant functions and for references to these constant functions in the main
dispatch. Accordingly, in the generated code, these names must have the same
variable ID as in the source program. Essentially, whenever a transformation
copies a name from the source program to the target program, the corresponding
ID must be copied as well and thus preserved. In contrast, names that are
synthesized by the transformation should have fresh variable IDs.

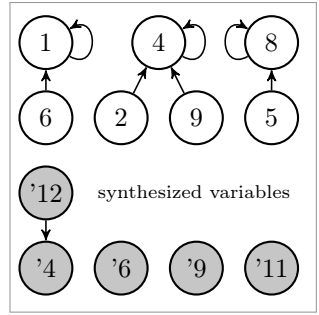
\begin{wrapfigure}{R}{.33\linewidth}
\vspace{-5ex}
\fboxgray{
    \begin{tikzpicture}[name graph]
      \node (dec-opened) at (1,2) {1}; 
      \node (dec-closed) at (2.5,2) {4};
      \node (dec-locked) at (4,2) {8};

      \node (ref-opened) at (1,1) {6};
      \node (ref-closed1) at (2,1) {2};
      \node (ref-closed2) at (3,1) {9};
      \node (ref-locked) at (4,1) {5};

      \node[synthesized, label=right:{\ \ \scriptsize synthesized variables}] (ref-opened-dispatch) at (1,0) {'12};

      \node[synthesized] (dec-opened-dispatch) at (1,-1) {'4}; 
      \node[synthesized] (dec-closed-dispatch) at (2,-1) {'6};
      \node[synthesized] (dec-locked-dispatch) at (3,-1) {'9};
      \node[synthesized] (dec-main) at (4,-1) {'11};

     \path (ref-opened) edge (dec-opened); 
     \path (ref-closed1) edge (dec-closed);
     \path (ref-closed2) edge (dec-closed);
     \path (ref-locked) edge (dec-locked);
     \path[min distance=3ex,out=-30,in=30] (dec-opened) edge (dec-opened);
     \path[min distance=3ex,out=-30,in=30] (dec-closed) edge (dec-closed);
     \path[min distance=3ex,out=210,in=150] (dec-locked) edge (dec-locked);

     \path (ref-opened-dispatch) edge (dec-opened-dispatch);
    \end{tikzpicture}
}
\caption{Names of compiled state machine of Figure~\ref{fig:door-simple}.}
  \label{fig:door-simple-graph}
\vspace{-5ex}
\end{wrapfigure}

For example, Figure~\ref{fig:door-simple-graph} shows the name graph of the
compiled state machine (we left out nodes of function parameters
\lstinline!event! and \lstinline!state! for clarity). We use line numbers from
the source program as variable IDs for reused variables, and \emph{ticked} line
numbers of the target program as variable IDs for synthesized variables. In
addition, we depict nodes of synthesized variables with a darker background
color. We have cycles in the name graph for source nodes 1, 4, and 8 because the
transformation duplicated the names at these labels to generate constant
functions and references to these constant functions.

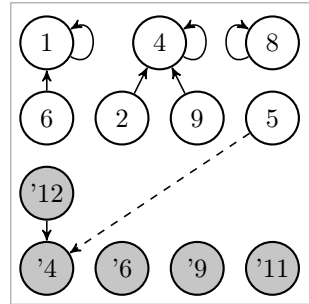
\begin{wrapfigure}{R}{.52\linewidth}
\vspace{-5ex}
\fboxgray{
    \begin{tikzpicture}[name graph]
      \node (dec-opened) at (1,2) {1}; 
      \node (dec-closed) at (2.5,2) {4};
      \node (dec-locked) at (4,2) {8};

      \node (ref-opened) at (1,1) {6};
      \node (ref-closed1) at (2,1) {2};
      \node (ref-closed2) at (3,1) {9};
      \node (ref-locked) at (4,1) {5};

      \node[synthesized] (ref-opened-dispatch) at (1,0) {'12};

      \node[synthesized] (dec-opened-dispatch) at (1,-1) {'4}; 
      \node[synthesized] (dec-closed-dispatch) at (2,-1) {'6};
      \node[synthesized] (dec-locked-dispatch) at (3,-1) {'9};
      \node[synthesized] (dec-main) at (4,-1) {'11};

     \path (ref-opened) edge (dec-opened); 
     \path (ref-closed1) edge (dec-closed);
     \path (ref-closed2) edge (dec-closed);
     \path[badref] (ref-locked) edge (dec-opened-dispatch);
     \path[min distance=3ex,out=-30,in=30] (dec-opened) edge (dec-opened);
     \path[min distance=3ex,out=-30,in=30] (dec-closed) edge (dec-closed);
     \path[min distance=3ex,out=210,in=150] (dec-locked) edge (dec-locked);

     \path (ref-opened-dispatch) edge (dec-opened-dispatch);
    \end{tikzpicture}
}
\caption{Variable capture (dashed arrow) in the code of Figure~\ref{fig:door-simple-renamed}.}
  \label{fig:door-simple-renamed-graph}
\vspace{-5ex}
\end{wrapfigure}

One important property of the name graph in Figure~\ref{fig:door-simple-graph}
is that the source nodes are disconnected from the synthesized nodes, and all
references from the original name graph in
Figure~\ref{fig:door-state-machine-graph} have been preserved. In contrast,
consider the name graph in Figure~\ref{fig:door-simple-renamed-graph} that
displays result of compilation after renaming state \lstinline!locked! to
\lstinline!opened-dispatch! as in Figure~\ref{fig:door-simple-renamed}. The graph illustrates that a source
variable has been captured (dashed arrow) during compilation: The variable at line~5 of the
source program was intended to point to the state declared at line~8, but after
compilation it points to the dispatch function at line~4 of the synthesized
program.

Our solution identifies variable capture by comparing the original name graph of
the whole program
with the name graph of the generated code. Function~\badBindings in
Figure~\ref{fig:fixCapture} computes the set of edges that witness variable
capture. In the state-machine example, \badBindings finds only one edge
$(\textnormal{5} \mapsto \textnormal{'4})$ as part of
\lstinline!notPresrvRef1!. We discuss the precise definition of variable capture
in the subsequent section.

If there are witnesses of variable capture, our solution computes a variable
renaming that has two properties. First, for each witness of variable capture,
the renaming renames the capturing variable to eliminate the witness. Second,
the renaming ensures that intentional references to the capturing variable are
renamed as well. This can be difficult because the name graph of the generated
code is inaccurate due to variable capture. Therefore, our solution
conservatively approximates the set of potential references by including all
synthesized variables of the same name. Function~\compRenamings in
Figure~\ref{fig:fixCapture} computes the renaming as a function from a variable
ID to the variable's fresh name, computed by \lstinline!gensym!. For the example, we get $\pisrc = \emptyset$
because $\textnormal{'4} \not\in V_s$ and $\pisyn = \{\textnormal{'4} \mapsto
\text{\lstinline{"opened-dispatch-0"}}, \textnormal{'12} \mapsto
\text{\lstinline{"opened-dispatch-0"}}\}$ because
$\nameat{\textnormal{'4}}{\textsf{t}} = \nameat{\textnormal{'12}}{\textsf{t}}$.
Function~\rename in Figure~\ref{fig:fixCapture} visits all nodes in a syntax
tree (represented as s-expression) and applies the renaming $\pi$ to variables with the corresponding IDs. For
the example, the renaming yields a capture-free program with the same name graph
as shown in Figure~\ref{fig:door-simple-graph}.

Function~\fixCapture in Figure~\ref{fig:fixCapture} brings it all together and
is the main entry point of our solution. It takes the name graph of the source
program and the generated target program as input. First, it computes the name
graph of the target program using the function $\resolve{T}$ that we assume to
provide name resolution for the target language $T$. \fixCapture then calls
\badBindings to identify variable capture. If \badBindings finds no capturing
edges, \fixCapture returns the generated program unchanged. Otherwise,
\fixCapture calls \compRenamings and \rename to compute and apply the renaming
that eliminates the witnessed variable capture. Since the name graph
\lstinline!G$_t$! of \lstinline!t! may be inaccurate due to variable capture,
\fixCapture recursively calls itself to repeat the search for and potential repair
of variable capture. Note that \fixCapture applies a closed-world assumption to
infer that all unbound variables are indeed free, and thus can be renamed at will.

\begin{figure}[tp]
\begin{lstlisting}[language=Pseudo,style=figureframe]
$\textnormal{Syntactic conventions:}$
˚˚x$^v$˚˚˚˚˚˚˚˚$\textnormal{variable \textsf{x} labeled with variable ID \textit{v}}$
˚˚$\nameat{v}{p} = x$˚˚˚$\textnormal{name \textit{x} that occurs in program \textit{p} at variable ID \textit{v}}$

$\badBindings$(($V_s$, $\rho_s$), ($V_t$, $\rho_t$)) = {
˚˚notPresrvRef1 = {($v$ $\mapsto$ $\rho_t$($v$)) | $v$ $\in$ $\dom$($\rho_t$), $v$ $\in$ $V_s$, $v$ $\in$ $\dom$($\rho_s$), $\rho_s$($v$) $\not=$ $\rho_t$($v$)};
˚˚notPresrvRef2 = {($v$ $\mapsto$ $\rho_t$($v$)) | $v$ $\in$ $\dom$($\rho_t$), $v$ $\in$ $V_s$, $v$ $\not\in$ $\dom$($\rho_s$), $v$ $\not=$ $\rho_t$($v$)};
˚˚notPresrvDef ˚= {($v$ $\mapsto$ $\rho_t$($v$)) | $v$ $\in$ $\dom$($\rho_t$), $v$ $\not\in$ $V_s$, $\rho_t$($v$) $\in$ $V_s$};
˚˚return notPresrvRef1 $\cup$ notPresrvRef2 $\cup$ notPresrvDef;
}

$\compRenamings$(($V_s$, $\rho_s$), ($V_t$, $\rho_t$), t, capture) = {
˚˚$\pisrc$ = $\emptyset$;
˚˚$\pisyn$ = $\emptyset$;
˚˚foreach $\vdef$ in $\codom$(capture) {
˚˚˚˚usedNames = $\{\nameat{v}{\textsf{t}} | v \in V_t\}$ $\cup$ $\codom$($\pisrc$)  $\cup$ $\codom$($\pisyn$)
˚˚˚˚fresh = gensym($\nameat{\vdef}{t}$, usedNames);
˚˚˚˚if ($\vdef$ $\in$ $V_s$ $\wedge$ $\vdef$ $\not\in$ $\pisrc$)
˚˚˚˚˚˚$\pisrc$ = $\pisrc$ $\cup$ {($\vdef$ $\mapsto$ fresh)} $\cup$ {($\vref$ $\mapsto$ fresh) | $\vref$ $\in$ $\dom$($\rho_s$), $\rho_s$($\vref$) = $\vdef$};
˚˚˚˚if ($\vdef$ $\not\in$ $V_s$ $\wedge$ $\vdef$ $\not\in$ $\pisyn$)
˚˚˚˚˚˚$\pisyn$ = $\pisyn$ $\cup$ {($v$ $\mapsto$ fresh) | $v$ $\in$ $V_t$ $\setminus$ $V_s$, $\nameat{v}{\textsf{t}}$ = $\nameat{\vdef}{\textsf{t}}$};
˚˚}
˚˚return ($\pisrc$, $\pisyn$);
}

(@*\rename\!\!*@)(t, $\pi$) = {
˚˚return t match {
˚˚˚˚case x$^v$ if v $\in$ $\dom$($\pi$)  =>  $\pi$(v)$^v$
˚˚˚˚case x$^v$˚˚˚˚˚˚˚˚˚˚˚˚˚$\hskip.35em$=>  x$^v$
˚˚˚˚case c  ˚˚˚˚˚˚˚˚˚˚˚˚˚$\hskip.175em$=>  c
˚˚˚˚case (t$_1$$ \dots $t$_n$)˚˚˚˚˚˚˚˚$\hskip.2em$=>  ($\rename$(t$_1$, $\pi$) $\ldots$ $\rename$(t$_n$, $\pi$));
˚˚}
}

$\fixCapture$(G$_s$, t) = {
˚˚G$_t$ = $\resolve{T}$(t);

˚˚capture = $\badBindings$(G$_s$, G$_t$);
˚˚if (capture == $\emptyset$) return t;

˚˚($\pisrc$, $\pisyn$) = $\compRenamings$(G$_s$, G$_t$, t, capture);
˚˚t' = $\rename$(t, $\pisrc$ $\cup$ $\pisyn$);
˚˚return $\fixCapture$(G$_s$, t');
}
\end{lstlisting}
  \caption{Definition of \fixCapture that guarantees capture-avoidance.}
  \label{fig:fixCapture}
\end{figure}

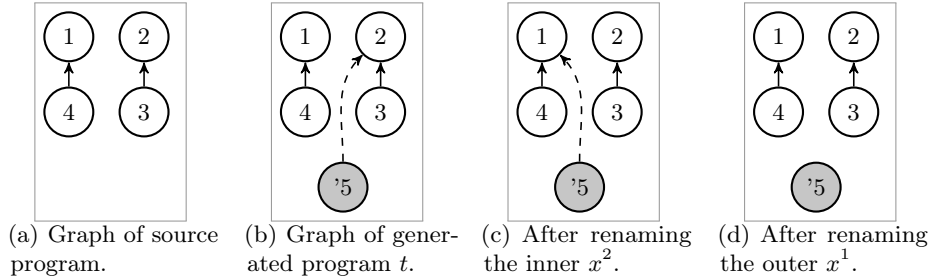
\begin{figure}[t]
  \def\mywidth{.2\linewidth}
  \begin{subfigure}
    \begin{minipage}{\mywidth}
      \begin{center}
        \fboxgray{
          \begin{tikzpicture}[name graph]
            \node (dec-outer) at (1,2) {1}; \node (dec-inner) at (2,2) {2};

            \node (ref-outer) at (1,1) {4}; \node (ref-inner) at (2,1) {3};

            \node[color=white] (invisible) at (1,0) {}; 

            \path (ref-outer) edge (dec-outer);
            \path (ref-inner) edge (dec-inner);
          \end{tikzpicture}
        }
      \end{center}
    \end{minipage}
    \caption{Graph of source program.}
  \end{subfigure}
\hfill
\begin{subfigure}
  \begin{minipage}{\mywidth}
    \begin{center}
      \fboxgray{
        \begin{tikzpicture}[name graph]
          \node (dec-outer) at (1,2) {1}; \node (dec-inner) at (2,2) {2};

          \node (ref-outer) at (1,1) {4}; \node (ref-inner) at (2,1) {3};

          \node[synthesized] (ref-syn) at (1.5,0) {'5};

          \path (ref-outer) edge (dec-outer); 
          \path (ref-inner) edge (dec-inner); 
          \path[badref,out=90,in=225] (ref-syn) edge (dec-inner);
        \end{tikzpicture}
      }
    \end{center}
  \end{minipage}
  \caption{Graph of generated program $t$.}
\end{subfigure}
\hfill
\begin{subfigure}
  \begin{minipage}{\mywidth}
    \begin{center}
      \fboxgray{
        \begin{tikzpicture}[name graph]
          \node (dec-outer) at (1,2) {1}; \node (dec-inner) at (2,2) {2};

          \node (ref-outer) at (1,1) {4}; \node (ref-inner) at (2,1) {3};

          \node[synthesized] (ref-syn) at (1.5,0) {'5};

          \path (ref-outer) edge (dec-outer); 
          \path (ref-inner) edge (dec-inner); 
          \path[badref,out=90,in=315] (ref-syn) edge (dec-outer);
        \end{tikzpicture}
      }
    \end{center}
  \end{minipage}
  \caption{After renaming the inner $x^2$.}
\end{subfigure}
\hfill
\begin{subfigure}
  \begin{minipage}{\mywidth}
    \begin{center}
      \fboxgray{
        \begin{tikzpicture}[name graph]
          \node (dec-outer) at (1,2) {1}; \node (dec-inner) at (2,2) {2};

          \node (ref-outer) at (1,1) {4}; \node (ref-inner) at (2,1) {3};

          \node[synthesized] (ref-syn) at (1.5,0) {'5};

          \path (ref-outer) edge (dec-outer); 
          \path (ref-inner) edge (dec-inner);
        \end{tikzpicture}
      }
    \end{center}
  \end{minipage}
  \caption{After renaming the outer $x^1$.}
\end{subfigure}
  \caption{Name graphs during execution of \fixCapture for $t = \lambda\,x^1.\, (\lambda\,x^2.\, x^3\, x^{\text{'}5})\, x^4$.}
  \label{fig:lambda-recursion-graphs}
\end{figure}

In the following, we present examples that illustrate two design choices of
\fixCapture that may be somewhat unintuitive: Why are multiple rounds of
renaming required, and why do we rename all synthesized variables of the same
name. For the former property, consider the lambda expression %
$t = \lambda\,x^1.\, (\lambda\,x^2.\, x^3\, x^{\text{'}5})\, x^4$, where we use
superscripts to annotate variable IDs and ticked IDs for synthesized variables.
The first graph in Figure~\ref{fig:lambda-recursion-graphs} shows the original
binding structure of the hypothetical source program that $t$ is generated from.
The second graph shows the binding structure of $t$. The synthesized variable
$x^{\text{'}5}$ is captured by the binding of $x^2$, which is illegal due to
\lstinline!notPresrvDef! in \badBindings. Accordingly, \compRenamings initiates
a renaming of $x^2$, also renaming $x^3$ to preserve the source reference. This
yields expression $t' = \lambda\,x^1.\, (\lambda\,\alpha^2.\, \alpha^3\,
x^{\text{'}5})\, x^4$ with binding structure as shown in the third graph.
Indeed, $x^2$ no longer captures $x^{\text{'}5}$. However, now $x^1$ captures
$x^{\text{'}5}$. Thus, by renaming $x^1$ and its reference $x^4$, we get $t'' =
\lambda\,\beta^1.\, (\lambda\,\alpha^2.\, \alpha^3\, x^{\text{'}5})\, \beta^4$
with capture-free binding structure as shown in the last graph. The iterative
renaming was necessary because the name graph of $t$ did not indicate that
$x^{\text{'}5}$ is eventually captured by $x^1$. We could have preemptively
renamed $x^1$ together with $x^2$, but this contradicts our goal for minimal
invasiveness.

To illustrate why \fixCapture renames all synthesized variables of the same
name, consider the expression %
$t = \lambda\,x^{\text{'}3}.\, x^1 (\lambda\,x^2.\, x^{\text{'}4})$ in which
$x^{\text{'}3}$ captures $x^{\text{'}1}$ and $x^2$ captures $x^{\text{'}4}$.
Thus, \fixCapture needs to
rename $x^{\text{'}3}$ and $x^2$. Because $x^{\text{'}3}$ and $x^{\text{'}4}$
are both synthesized and have the same name, renaming of $x^{\text{'}3}$ entails
the renaming of $x^{\text{'}4}$ even though they are unrelated in the name graph
of $t$. Thus, \fixCapture yields the correct result $t' =
\lambda\,\alpha^{\text{'}3}.\, x^1 (\lambda\,\beta^2.\, \alpha^{\text{'}4})$. To
see why $x^{\text{'}3}$ should bind $x^{\text{'}4}$, consider what happens had
the source program consistently used $y$ in place of $x$: %
$t_2 = \lambda\,x^{\text{'}3}.\, y^1 (\lambda\,y^2.\, x^{\text{'}4})$. This
program has no variable capture and is returned unchanged by \fixCapture. Since
we want the result of \fixCapture to be invariant under consistent renamings of
the source variables, $x^{\text{'}3}$ must bind $x^{\text{'}4}$ in both $t$ and
$t_2$. By renaming all synthesized variables of the same name, \fixCapture
ensures that no potential variable reference is truncated.

Both of the above examples also illustrate another point: \fixCapture does not
guarantee valid name binding with respect to the target language. The final
result in both examples contains a free variable. Instead, \fixCapture
guarantees that there is no variable capture. We state and verify the precise
properties of \fixCapture in the next section.

\section{Termination, correctness, and an equivalence theory}
\label{sec:formalization}

Our solution \fixCapture iteratively eliminates variable capture in a
fixed-point computation. In this section we show three important properties of
\fixCapture: \fixCapture terminates, \fixCapture eliminates variable capture,
and \fixCapture yields $\alpha$-equivalent outputs for inputs that are equal up
to consistent (but possibly capturing) variable renaming.

We represent programs as s-expressions with constant symbols $c$, labeled
variable names $x^v$, and compound terms $(t_1 \dots t_n)$. We shall
frequently require two programs to be equal up to unconditional renaming:
\begin{definition}
  \textnormal{Two programs are \emph{label-equivalent} $p_1
   \equivL p_2$ iff they are equal up to variable names:}
{
\addtolength\abovedisplayskip{-3ex}
\addtolength\belowdisplayskip{-3ex}
\[
\begin{array}{rcl@{\hskip2em}l}
  c_1 & \equivL & c_2 & \mathrm{if}\ c_1 = c_2 \\
  x_1^{v_1} & \equivL & x_2^{v_2} & \mathrm{if}\ v_1 = v_2 \\
  (t_1 \ldots t_n) & \equivL & (t'_1 \ldots t'_n) & \mathrm{if}\ t_i \equivL t'_i\ \ \forall\, 1 \leq i \leq n
\end{array}
\]
}
\end{definition}
To simplify our formalization, we do not consider bijective relabeling
functions and assume label-equivalence instead. As first metatheoretical result
we state that \fixCapture terminates.%
\iftechreport
\footnote{Proofs of all theorems and additional lemmas appear in Appendix~\ref{app:proofs}.}
\else 
\footnote{Proofs of theorems and additional lemmas appear in a technical report~\cite{ErdwegSD14a}}
\fi

\newcommand{\lemmaOfTermination}{%
\begin{theorem}\label{thm:termination}
  For any name graph $G_s$ and any program~$t$,
  $\fixCapture(G_s, t)$ terminates in finitely many steps.
\end{theorem}}
\newcommand{\proofOfTermination}{
 \proof{The depth of the recursion of \fixCapture is bound by the number of
  variable declarations in $t$. Each variable declaration $\vdef$ can at most occur once
  in the result of \badBindings because it is immediately renamed to
  a fresh name. The renamed variable declaration cannot occur in \badBindings
  again because (i)~if  $\vdef \in V_s$, then only references $\vref \in V_s$
  with $\rho(\vref) = \vdef$ share the fresh name and
  (ii)~if $\vdef \not\in V_s$, then only references $\vref \in V_t \setminus V_s$
  share the fresh name but $\vref \in \dom(\badBindings)$ entails
  $\badBindings(\vref) \in V_s$. Hence \fixCapture terminates after at most all variable declarations in $t$
  have been renamed once.
 \qed}
}
\lemmaOfTermination

\subsection{Assumptions on name resolution}

We present our framework for capture-avoiding transformations independent of any
concrete source and target languages. Since our technique works on top of name
graphs, we require functions \resolve{L} that compute the name
graph of a program of some language $L$ by name analysis. However, instead of
requiring a specific form of name analysis, we specify minimal requirements on
the behavior of \resolve{L} that suffice to show our technique is sound. The
first assumption states that name analysis must produce a name graph.

\begin{assumption}
 \label{ass:resolve:return-name-graph}
 Given a program $p$, $\resolve{L}(p)$ yields the name graph $G = (V,
 \rho)$ of $p$ according to Definition~\ref{def:name-graph}.
\end{assumption}

\noindent The second assumption requires \resolve{L} to behave
deterministically. First, given two programs $p_1$ and $p_2$ that are equal up
to variable names, names that are references in $p_1$ must be references in
$p_2$ if the declaration is available (but it can refer to another
declaration). Second, given a reference with two potential declarations in $p_1$
and $p_2$, $\resolve{L}$ must deterministically choose one of them.

\begin{assumption}
  Let $p_1 \equivL p_2$ be label-equivalent with name graphs %
  $\resolve{L}(p_1) = (V, \rho_1)$ and %
  $\resolve{L}(p_2) = (V, \rho_2)$. %
\vspace{-1ex}
  \begin{enumerate}
  \item[(i)]\label{ass:resolve:also-ref} If $\rho_1(\vref) = \vdef$ and $\nameat{\vref}{p_2} =
    \nameat{\vdef}{p_2}$, then $\vref \in \dom(\rho_2)$. %
  \item[(ii)]\label{ass:resolve:deterministic}
    If $\rho_1(\vref) = \vdef$, $\rho_2(\vref)= \vdef'$, %
    $\nameat{\vdef}{p_1} = \nameat{\vdef'}{p_1}$, and $\nameat{\vdef}{p_2} =
    \nameat{\vdef'}{p_2}$, then $\vdef = \vdef'$.
  \end{enumerate}
\end{assumption}

\noindent In addition to these assumptions, we require that the name graph
$(V,\rho)$ of the original source program satisfies $\dom(\rho) \cap
\codom(\rho) = \emptyset$. We call such graphs \emph{bipartite name
graphs}. Note that $\resolve{L}$ often does not produce bipartite name graphs
for generated code due to name copying as in
Figure~\ref{fig:door-simple-graph}. We believe our requirements are modest and
readily satisfied by name analyses of most languages.



\subsection{\fixCapture eliminates variable capture}

We define the notion of capture-avoiding transformations in terms of the name
graph of the source and target programs, before we show that \fixCapture can
turn any transformation into a capture-avoiding one.

\begin{definition}
  \textnormal{ A transformation $f: S \to T$ is \emph{capture-avoiding} if for
    all $s \in S$ with $\resolve{S}(s) = (V_s, \rho_s)$ and $t = f(s)$ with
    $\resolve{T}(t) = (V_t, \rho_t)$:
  \begin{enumerate}
  \item Preservation of reference intent:
    For all $v \in \dom(\rho_t)$ with $v \in V_s$,
    \begin{enumerate}
    \item[(i)] if $v \in \dom(\rho_s)$, then $\rho_s(v) = \rho_t(v)$,
    \item[(ii)] if $v \not\in \dom(\rho_s)$, then $v = \rho_t(v)$.
    \end{enumerate}
  \item Preservation of declaration extent:
    For all $v \in \dom(\rho_t)$, if $v \not\in V_s$, then $\rho_t(v) \not\in V_s$.
%
%
%
  \end{enumerate}
}
\end{definition}
The first condition states that a capture-avoiding transformation must preserve
references of the source program. That is, if a variable $v$
occurs in the target program and this reference was bound in the source program,
then the target program must provide the same binding for $v$. That is, the
transformation must preserve the reference intent of the source program's
author.

If the source program does not contain $v$ as a bound variable (but maybe as
a declaration), $v$ can only refer to itself in the target program. We
specifically admit such self-references to allow transformations to duplicate
names of source-program declarations in order to introduce additional
delegation. For example, our compiler for state machines illustrated in
Figure~\ref{fig:door-state-machine} uses names of state declarations to generate
constant functions and references to these functions. Note that we also admit
duplication of reference names, each of which has the same variable ID and thus
must refer to the original declaration.

The second condition states that a capture-avoiding transformation must keep
synthesized variable references separate from variables declared in the source
program. We consider all variables of the source program $V_s$ to be original
and all variables of the target program that do not come from the source program
$(V_t \setminus V_s)$ to be synthesized. This condition prevents synthesized
variable references to be captured by original variable declarations, that is,
synthesized variables can only be bound by synthesized declarations.

Function \badBindings in Figure~\ref{fig:fixCapture} implements the test for
capture avoidance and collects witnesses in case of variable capture. Since
\fixCapture only terminates when \badBindings fails to find variable capture,
the correctness of \fixCapture follows from its termination.

\newcommand{\lemmaOfCaptureAvoidance}{%
\begin{theorem}[Capture avoidance] \label{thm:capture-avoidance}
 \textnormal{
  Given a transformation $f : S \to T$, \linebreak[4]\fixCapture yields a capture-avoiding
  transformation $\lambda s.\,\fixCapture(\resolve{S}(s), f(s))$.
 }
\end{theorem}}
\newcommand{\proofOfCaptureAvoidance}{
 \proof{When \fixCapture terminates, $\badBindings = \emptyset$ and thus all
   reference intent and declaration extent is preserved from the name graph of
   $s$ to the name graph of the resulting program.
 \qed}
}
\lemmaOfCaptureAvoidance

\subsection{Definitions of $\alpha$-equivalence and sub-$\alpha$-equivalence}

It is not enough to ensure that \fixCapture eliminates variable capture,
because, for example, a function that returns the empty program would satisfy
this property. To ensure the usefulness of \fixCapture, we need to show that,
given two programs that are equal up to possibly capturing renaming, it produces
$\alpha$-equivalent programs (and not just any programs).
Two programs are $\alpha$-equivalent if they are equal up to non-capturing
renaming, that is, if they have the same syntactic structure and binding structure.

\begin{definition}\label{def:alpha-equivalence}
  \textnormal{Two programs $p_1$ and $p_2$ with name graphs $\resolve{L}(p_1) =
    (V_1, \rho_1)$ and $\resolve{L}(p_2) = (V_2, \rho_2)$ are
    \emph{$\alpha$-equivalent} $p_1 \equivA p_2$ iff $p_1 \equivL p_2$ and $\rho_1 = \rho_2$. }


\end{definition}
Note that $p_1 \equivL p_2$ entails $V_1 = V_2$. As expected, our definition of
$\alpha$-equivalence is independent of the concrete names that occur in the
programs. The following examples illustrate our definition of
$\alpha$-equivalence.
\begin{center}
  \begin{tabular}{r@{\hskip.3em}l@{\hskip.8em}r@{\hskip.5em}l}
    \toprule
    \multicolumn{2}{l}{Program} & \multicolumn{2}{l}{Name graph} \\\midrule
    $p_1 =$  & $\lambda\, x^1.\ (\lambda\, y^3.\ y^4\ y^5)\ x^2$   & $G_1 =$ & $(\{1,2,3,4,5\}, \{(2 \mapsto 1),(4 \mapsto 3),(5 \mapsto 3)\}) $ \\
    $p_2 =$  & $\lambda\, x^1.\ (\lambda\, x^3.\ x^4\ x^5)\ x^2$   & $G_2 =$ & $(\{1,2,3,4,5\}, \{(2 \mapsto 1),(4 \mapsto 3),(5 \mapsto 3)\}) $ \\
    $p_3 =$  & $\lambda\, x^1.\ (\lambda\, y^3.\ x^4 + y^5)\ x^2$  & $G_3 =$ & $(\{1,2,3,4,5\}, \{(2 \mapsto 1),(4 \mapsto 1),(5 \mapsto 3)\}) $ \\
    $p_4 =$  & $\lambda\, x^1.\ (\lambda\, x^3.\ x^4 + x^5)\ x^2$  & $G_4 =$ & $(\{1,2,3,4,5\}, \{(2 \mapsto 1),(4 \mapsto 3),(5 \mapsto 3)\}) $ \\
    \bottomrule
  \end{tabular}
\end{center}
Our definition correctly identifies $p_1 \equivA p_2$, because they are
label-equivalent and have the same name graphs. Indeed, $p_2$ can be derived
from $p_1$ by consistently renaming all occurrences of the bound variable $y$ to
$x$. In contrast, $p_3 \not\equivA p_4$ because the binding structure differs:
$x^4$ is bound to $x^1$ in $p_3$, but to $x^3$ in $p_4$. All other combinations
of above programs (modulo symmetry of $\equivA$) are not $\alpha$-equivalent
because they fail the required label-equivalence. In particular, $p_2
\not\equivA p_4$ in spite of having the same binding structure.

To relate programs that are equal up to possibly capturing renaming, we propose
the following notion of sub-$\alpha$-equivalence.

\begin{definition}
  \textnormal{Two programs are \emph{sub-$\alpha$-equivalent} $p_1 \equivSA{G} p_2$
    under a name graph $G = (V, \rho)$ iff $p_1 \equivL p_2$ and, given $V_p$
    is the set of labels in $p_1$ and $p_2$,
$
\begin{array}{rl@{\hskip1em}l}
  \text{(i)}  & \text{for all } \vref, \vdef \in V_p \cap V \text{ with } \rho(\vref) = \vdef, & %
    \nameat{\vref}{p_1} = \nameat{\vdef}{p_1} \Leftrightarrow \nameat{\vref}{p_2} = \nameat{\vdef}{p_2} \\
  \text{(ii)} & \text{for all } \vref, \vdef \in V_p \setminus V, & %
    \nameat{\vref}{p_1} = \nameat{\vdef}{p_1} \Leftrightarrow \nameat{\vref}{p_2} = \nameat{\vdef}{p_2} \\
\end{array}
$
}
\end{definition}
Sub-$\alpha$-equivalence compares two programs based on the actual names
occurring in them, and not based on the binding structure. The relation is
parameterized over a name graph $G$. The first condition states that for each
binding in this graph, $p_1$ and $p_2$ need to agree on whether reference and
declaration share the same name or not. Even if the reference and declaration
have the same name, it does not imply that there is a corresponding binding in
either $p_1$ or $p_2$, because another declaration can also have this name and
capture the reference. The second condition states that for all variables not in
$G$, $p_1$ and $p_2$ need to agree on which variable occurrences share names.
To illustrate sub-$\alpha$-equivalence, let us consider $G = (\{1,2,3\}, \{(2
\mapsto 1),(3 \mapsto 1)\})$ and the following programs:

\vskip.5\baselineskip
\noindent
\begin{tabular}{r@{\hskip2em}r@{\hskip.5em}l@{\hskip3em}r@{\hskip.5em}l}
    \toprule
\multirow{2}{*}{$[p_1]_{\equivSA{G}}$} &    
     $p_1 =$  & $\lambda\, x^1.\ (\lambda\, y^{\tick 4}.\ x^3 + y^{\tick 5})\ x^2$ 
   & $p_2 =$  & $\lambda\, z^1.\ (\lambda\, y^{\tick 4}.\ z^3 + y^{\tick 5})\ z^2$ \\
   & $p_3 =$  & $\lambda\, x^1.\ (\lambda\, z^{\tick 4}.\ x^3 + z^{\tick 5})\ x^2$
   & $p_4 =$  & $\lambda\, z^1.\ (\lambda\, z^{\tick 4}.\ z^3 + z^{\tick 5})\ z^2$ \\
\midrule
\multirow{2}{*}{$\neg [p_1]_{\equivSA{G}}$} &
     $p_5 =$  & $\lambda\, \highlight{z}^1.\ (\lambda\, y^{\tick 4}.\ x^3 + y^{\tick 5})\ x^2$ 
   & $p_6 =$  & $\lambda\, x^1.\ (\lambda\, y^{\tick 4}.\ \highlight{z}^3 + y^{\tick 5})\ x^2$ \\
   & $p_7 =$  & $\lambda\, x^1.\ (\lambda\, \highlight{z}^{\tick 4}.\ x^3 + y^{\tick 5})\ x^2$
   & $p_8 =$  & $\lambda\, x^1.\ (\lambda\, y^{\tick 4}.\ x^3 + \highlight{z}^{\tick 5})\ x^2$ \\
    \bottomrule
  \end{tabular}
\vskip.5\baselineskip

\noindent %
The first four programs are sub-$\alpha$-equivalent to $p_1$ under $G$. We have
$p_1 \equivSA{G} p_2$ because they agree on the name sharing at variable IDs
$1$, $2$, and $3$, which is required because of the bindings in $G$, and on the
name sharing at variable IDs $\tick 4$ and $\tick 5$, which is required because
these IDs are not in $G$. Similar analysis shows $p_1 \equivSA{G} p_3$ and $p_1
\equivSA{G} p_4$.
Programs $p_5$ through $p_8$ are examples that are not sub-$\alpha$-equivalent
to $p_1$ under $G$. For $p_5$ and $p_6$ the first condition of
sub-$\alpha$-equivalence fails because there is no agreement on the name sharing
at $1$ and $3$. For $p_7$ and $p_8$ the second condition fails because there is
no agreement on the name sharing at $\tick 4$ and $\tick 5$.

Note that $p_1 \equivSA{G} p_4$ illustrates that sub-$\alpha$-equivalence is
weaker than $\alpha$-equivalence because $p_1 \not\equivA p_4$. In the following
subsection we use sub-$\alpha$-equivalence to characterize programs that
\fixCapture can repair to $\alpha$-equivalent programs. 

\newcommand{\lemmaOfSubAlphaEquiv}{%
\begin{lemma}
 For any graph $G$, sub-$\alpha$-equivalence under $G$ is an equivalence relation,
 that is, it is reflexive, symmetric, and transitive.
\end{lemma}}
\newcommand{\proofOfSubAlphaEquiv}{
 \proof{Follows directly from the definition of sub-$\alpha$-equivalence and the fact that $\equivL$ is an equivalence relation. \qed}
}

\subsection{An equivalence theory for \fixCapture}

We now turn to one of the main results of our metatheory: Function \fixCapture
is noninvasive, preserves sub-$\alpha$-equivalence, and is invariant under
consistent (but possibly capturing) renaming of original and synthesized
variables, as specified by sub-$\alpha$-equivalence. %

For capture-free programs, \fixCapture yields the input program unchanged, that is, \fixCapture is noninvasive:
\newcommand{\lemmaOfFixCaptureBaseId}{%
\begin{theorem}\label{thm:fixCapture-base-id}
 \textnormal{
  For any name graph $G_s = (V_s, \rho_s)$ and any program $t$ with $\badBindings(G_s, \resolve{T}(t))
  = \emptyset$, $\fixCapture(G_s, t) = t$.
 }
\end{theorem}}
\newcommand{\proofOfFixCaptureBaseId}{\proof{By definition of \fixCapture. \qed}}
\lemmaOfFixCaptureBaseId

\newcommand{\lemmaOfRenamingPreservesStructure}{%
\begin{lemma}\label{lem:renaming-preserves-structure}
 For any renaming $\pi$ and program $t$, $\rename(t, \pi) \equivL t$.
\end{lemma}}
\newcommand{\proofOfRenamingPreservesStructure}{\proof{By induction on the structure of $t$ \qed}}


\newcommand{\lemmaOfFixCaptureBaseAlpha}{%
\begin{lemma}\label{lem:fixCapture-base-alpha}
 For any name graph $G_s = (V_s, \rho_s)$ and sub-$\alpha$-equivalent programs
 $t_1 \equivSA{G_s} \!t_2$ under name graph $G_s$, if $\badBindings(G_s,
 \resolve{T}(t_1)) = \emptyset$ and $\badBindings(G_s, \resolve{T}(t_2)) =
 \emptyset$, then $t_1 \equivA t_2$.
\end{lemma}}
\newcommand{\proofOfFixCaptureBaseAlpha}{
 \proof{Let $(V_i, \rho_i) = \resolve{T}(t_i)$ and \lstinline{capture}$_i$ $= \badBindings(G_s, \resolve{T}(t_i))$. By definition $t_1 \equivA t_2$
  if $\rho_1 = \rho_2$, which holds if $\dom(\rho_1) =
  \dom(\rho_2)$ and $\rho_1(v) = \rho_2(v)$ for all $v \in \dom(\rho_1)$. Let $v
  \in \dom(\rho_1)$ (analogously for $v \in \dom(\rho_2)$). We distinguish 3 cases:
  \vspace{-1.0ex} \begin{enumerate}
  \item If $v \in V_s$ and $v \in \dom(\rho_s)$, then $\rho_1(v) =
    \rho_s(v)$
    because otherwise $\rho_s(v) \not= \rho_1(v)$ entails $(v \mapsto
    \rho_1(v)) \in$ \lstinline{notPresrvRef1} $\subseteq$ \lstinline{capture}$_1$, contradicting
    \lstinline{capture}$_1$ $= \emptyset$.
    By Assumption~\ref{ass:resolve:return-name-graph}, $\nameat{v}{t_1} = \nameat{\rho_1(v)}{t_1}$, which implies
    $\nameat{v}{t_2} = \nameat{\rho_1(v)}{t_2}$ by the first condition of $t_1 \equivSA{G_s} \!t_2$.
    Thus by Assumption~\ref{ass:resolve:also-ref}-(i), $v \in \dom(\rho_2)$.
    Then $\rho_2(v) = \rho_s(v)$ because otherwise $(v \mapsto
    \rho_2(v)) \in$ \lstinline{notPresrvRef1} $\subseteq$ \lstinline!capture!$_2$, contradicting
    \lstinline{capture}$_2$ $= \emptyset$.
    By Assumption~\ref{ass:resolve:also-ref}-(ii), $\rho_1(v) = \rho_2(v)$.
  \item If $v \in V_s$ and $v \not\in \dom(\rho_s)$, then $\rho_1(v) = v$,
    because otherwise $(v \mapsto \rho_1(v)) \in$
    \lstinline!notPresrvRef2! $\subseteq$ \lstinline!capture!$_1$,
    contradicting \lstinline!capture!$_1$ $= \emptyset$.
    We trivially have $\nameat{v}{t_2} = \nameat{\rho_1(v)}{t_2}$ and
    thus by Assumption~\ref{ass:resolve:also-ref}-(i), $v \in \dom(\rho_2)$.  Then $\rho_2(v) = v$,
    by because otherwise $(v \mapsto \rho_2(v)) \in$ \lstinline!notPresrvRef2! $\subseteq$ \lstinline!capture!$_2$,
    contradicting \lstinline!capture!$_2$ $= \emptyset$.
    By Assumption~\ref{ass:resolve:deterministic}-(ii), $\rho_1(v) = v = \rho_2(v)$.
  \item If $v \not\in V_s$, then $\rho_1(v) \not\in V_s$ because otherwise $(v \mapsto \rho_1(v)) \in $ \lstinline!notPresrvDef! $\subseteq$ \lstinline!capture!$_1$,
    contradicting \lstinline!capture!$_1$ $= \emptyset$.
    By Assumption~\ref{ass:resolve:return-name-graph}, $\nameat{v}{t_1} = \nameat{\rho_1(v)}{t_1}$, which implies
    $\nameat{v}{t_2} = \nameat{\rho_1(v)}{t_2}$ by the second condition of $t_1 \equivSA{G_s} \!t_2$.
    By Assumption~\ref{ass:resolve:also-ref}-(i), $v \in
    \dom(\rho_2)$. We have $\rho_2(v) \not\in V_s$ because otherwise $(v \mapsto \rho_2(v)) \in $ \lstinline!notPresrvDef! $\subseteq$ \lstinline!capture!$_2$,
    contradicting \lstinline!capture!$_2$ $= \emptyset$. 
    By Assumption~\ref{ass:resolve:return-name-graph}, $\nameat{v}{t_1} = \nameat{\rho_1(v)}{t_1}$ and thus
    $\rho_1(v) = \rho_2(v)$ by Assumption~\ref{ass:resolve:deterministic}.
\qed
  \end{enumerate} %
 }
}

\newcommand{\lemmaOfFixCaptureStepPreservesSubAlpha}{%
\begin{lemma}\label{lem:fixCapture-step-preserves-sub-alpha}
 For any bipartite name graph $G_s = (V_s, \rho_s)$ and program $t$ with  
 \lstinline!capture! $= \badBindings(G_s, \resolve{T}(t)) \neq \emptyset$, renaming preserves
 sub-$\alpha$-equivalence $t \equivSA{G_s} \!\rename(t, \pisrc \cup \pisyn)$
 given $\pisrc$ and $\pisyn$ as in \fixCapture.
\end{lemma}}
\newcommand{\proofOfFixCaptureStepPreservesSubAlpha}{
 \proof{Let $(V_t, \rho_t) = \resolve{T}(t)$ and $t' = \rename(t, \pisrc \cup \pisyn)$.
  First we have $t \equivL t'$ by Lemma~\ref{lem:renaming-preserves-structure}.
  By the definition of \rename and since $\dom(\pisrc) \cap \dom(\pisyn) = \emptyset$,
  $\nameat{v}{t'}$ becomes either $\pisrc(v)$ if $v \in \dom(\pisrc)$, $\pisyn(v)$
  if $v \in dom(\pisyn)$, and remains unchanged otherwise. We separately show that both
  conditions of sub-$\alpha$-equivalence are satisfied:
  \vspace{-1.0ex} \begin{enumerate}
  \item For all $\vref, \vdef \in V_t \cap V_s$ with $\rho_s(\vref) = \vdef$ we have $\vref
    \notin \dom(\pisyn)$, $\vdef \notin \dom(\pisyn)$, $\vref \in \dom(\pisrc)
    \Leftrightarrow \vdef \in \dom(\pisrc)$ because $G_s$ is bipartite, and if $\vref \in \dom(\pisrc)$,
    then $\pisrc(\vref) = \pisrc(\vdef)$. Thus, $\nameat{\vref}{t} = \nameat{\vdef}{t} \Leftrightarrow \nameat{\vref}{t'} = \nameat{\vdef}{t'}$.
  \item For all $\vref, \vdef \in V_t \setminus V_s$ we have $\vref \not\in
    \dom(\pisrc)$ and $\vdef \not\in \dom(\pisrc)$. If $\nameat{\vref}{t} \neq
    \nameat{\vdef}{t}$, then $\nameat{\vref}{t'\,\!} \neq \nameat{\vdef}{t'\,\!}$
    because $\pisyn$ maps distinct names to distinct fresh names. %
    If instead $\nameat{\vref}{t} = \nameat{\vdef}{t}$, we
    have $\vref \in \dom(\pisyn) \Leftrightarrow \vdef \in \dom(\pisyn)$
    and if $\vref \in \dom(\pisyn)$, then $\pisyn(\vref) = \pisyn(\vdef)$.
    Thus, $\nameat{\vref}{t} = \nameat{\vdef}{t} \Leftrightarrow \nameat{\vref}{t'\,\!} = \nameat{\vdef}{t'\,\!}$.
 \qed
  \end{enumerate} \vspace{-1.0ex}}
}

\noindent Given a bipartite name graph of the source program, \fixCapture preserves sub-$\alpha$-equivalence:
\newcommand{\lemmaOfFixCapturePreservesSubAlpha}{%
\begin{theorem}\label{lem:fixCapture-preserves-sub-alpha}
 For any bipartite name graph $G_s = (V_s,\rho_s)$ and any program $t$, $\fixCapture(G_s, t) \equivSA{G_s} t$.
\end{theorem}}
\newcommand{\proofOfFixCapturePreservesSubAlpha}{
 \proof{By induction on $\fixCapture(G_s, t)$ using
  Theorem~\ref{thm:fixCapture-base-id} and
  Lemma~\ref{lem:fixCapture-step-preserves-sub-alpha}. \qed}
}
\lemmaOfFixCapturePreservesSubAlpha

\newcommand{\lemmaOfFixCaptureInnerInduction}{%
\begin{lemma}\label{lem:fixCapture-inner-induction}
 For any bipartite name graph $G_s = (V_s, \rho_s)$ and programs
 $t_1 \equivSA{G_s} \!t_2$, if $\badBindings(G_s, \resolve{T}(t_1))
 = \emptyset$, then $t_1 \equivA \fixCapture(G_s, t_2)$.
\end{lemma}}
\newcommand{\proofOfFixCaptureInnerInduction}{
 \proof{ By induction on $\fixCapture(G_s, t_2)$. Base case: $\badBindings(G_s,
  \resolve{T}(t_2))\linebreak[2] = \emptyset$ and $\fixCapture(G_s, t_2) = t_2$.
  Then $t_1 \equivA t_2$ by Lemma~\ref{lem:fixCapture-base-alpha}. Step
  case:\linebreak[4] $\badBindings(G_s, \resolve(t_2)) \neq \emptyset$ and
  $\fixCapture(G_s, t_2) = \fixCapture(G_s, t_2')$. Then $t_1 \equivSA{G_s} t_2'$ by
  Lemma~\ref{lem:fixCapture-step-preserves-sub-alpha}, and $t_1 \equivA
  \fixCapture(G_s, t_2')$ by the induction hypothesis.
 \qed}
}

\noindent Given a bipartite name graph of the source program, \fixCapture maps sub-$\alpha$-equivalent programs to $\alpha$-equivalent ones:
\newcommand{\lemmaOfFixCaptureSubAlphaInvariant}{%
\begin{theorem}\label{thm:fixCapture-sub-alpha-invariant}
 For any bipartite name graph $G_s = (V_s, \rho_s)$ and programs
 $t_1 \equivSA{G_s} \!t_2$, $\fixCapture(G_s, t_1)
 \equivA \fixCapture(G_s, t_2)$.
\end{theorem}}
\newcommand{\proofOfFixCaptureSubAlphaInvariant}{
 \proof{ By induction on $\fixCapture(G_s, t_1)$. Base case by
  Lemma~\ref{lem:fixCapture-inner-induction}. Step case: $\badBindings(G_s,
  \resolve(t_1)) \neq \emptyset$ and $\fixCapture(G_s, t_1) = \fixCapture(G_s, t_1')$.
  Then $t_1' \equivSA{G_s} t_1$ by Lemma~\ref{lem:fixCapture-step-preserves-sub-alpha} and $t_1' \equivSA{G_s} t_2$ by transitivity. Thus, $\fixCapture(G_s, t_1')
  \equivSA{G_s} \fixCapture(G_s, t_2)$ by the induction hypothesis.
 \qed}
}
\lemmaOfFixCaptureSubAlphaInvariant

\section{Hygienic transformations} \label{sec:hygienic-transformations}

In the previous section, we demonstrated that for any transformation $f : S
\to T$, \fixCapture provides a capture-avoiding transformation $\lambda\, s.
\fixCapture(G_s, f(s))$. However, for some transformations \fixCapture yields a
transformation that adheres to the stronger property of hygienic
transformations.

\begin{definition}
  \textnormal{A transformation $f : S \to T$ is \emph{hygienic} if it maps $\alpha$-equivalent
    source programs to $\alpha$-equivalent target programs: }
\[s_1 \equivA s_2  \quad\Longrightarrow\quad f(s_1) \equivA f(s_2). \]
\end{definition}
This definition of hygiene for transformations follows Herman's definition
of hygiene for syntax macros~\cite{Herman12}.

Transformations can inspect the names of variables
and can generate structurally different code for $\alpha$-equivalent
inputs. For example, a transformation may decide to produce
thread-safe accessors for variables with names prefixed by
\lstinline!sync_!. Accordingly, a consistent renaming from
\lstinline!sync_foo! to \lstinline!foo! in the source program leads
to generated programs that are not structurally equivalent, let alone
$\alpha$-equivalent.
However, there is an interesting class of transformations for which \fixCapture 
provides hygiene:
\begin{definition}
  \textnormal{A transformation $f : S \to T$ is \emph{sub-hygienic} if it maps
    $\alpha$-equivalent source programs $s_1 \equivA s_2$ to sub-$\alpha$-equivalent
    target programs $f(s_1) \equivSA{G_s} \!f(s_2)$ under the name graph $G_s$ of
    $s_1$ (or $s_2$).
}
\end{definition}
The class of sub-hygienic transformations includes some common transformation
schemes. First, it includes transformations that transform a source program
solely based on the program's structure but independent of the concrete variable
names occurring in it. In such transformations, synthesized variable names are
constant and the same for any source program. Second, for a source language
without name shadowing (such as state machines), sub-hygienic transformations
include those that derive synthesized variable names using an injective function
$g : \tstring \to \tstring$ over the corresponding source variable names. For
example, in Figure~\ref{fig:door-state-machine-simple}, we derived the name of a
dispatch function by appending \lstinline!-dispatch! to the
corresponding state name. In both cases \fixCapture eliminates all potential
variable capture and yields a fully hygienic transformation:

\newcommand{\lemmaOfNameFixHygiene}{%
\begin{theorem}\label{thm:name-fix-hygiene}
 For any sub-hygienic transformation $f : S \to T$, transformation $\lambda\,
 s. \fixCapture(G_s, f(s))$ is hygienic.
\end{theorem}}
\newcommand{\proofOfNameFixHygiene}{
 \proof{ For any $s_1 \equivA s_2$, $f(s_1) \equivSA{G_s} f(s_2)$ by the definition
  of sub-hygiene. Then $\fixCapture(G_s, f(s_1)) \equivA \fixCapture(G_s, f(s_2))$ by
  Theorem~\ref{thm:fixCapture-sub-alpha-invariant}.
 \qed }
}
\lemmaOfNameFixHygiene

\section{Case studies}
\label{SECT:casestudies}

To evaluate the applicability of capture-avoiding program transformation in practice, we
have successfully applied \fixCapture in three different scenarios:
\begin{itemize}
\item Optimization: Function inlining via substitution in a procedural language.
\item Desugaring of language extensions: Lambda lifting of local functions.
\item Code generation: Compilation of state machines and of Derric, an existing
  DSL for digital forensics, to Java.
\end{itemize}
We have implemented all case-studies in Rascal, a programming language and
environment for source code analysis and transformation~\cite{KlintSV09}. The
source code of our implementation and all case studies are available online:
\url{http://github.com/seba--/hygienic-transformations}.

\subsection{Preservation of variable IDs with string origins in Rascal}

As described in Section~\ref{sec:namefix-explanation}, a transformation must
preserve variable IDs of the source program when reusing these names in the
target program. While it is possible for a developer of a program transformation
to manually preserve variable IDs via copying, it is easier and safer if the
transformation engine does it automatically. We extended
Rascal to preserve variable IDs automatically via a new Rascal feature called
\emph{string origins}~\cite{ValderaSE14}. Every string value (captured by the \irascal{str} data
type) carries information about its origin. A string can either originate from a
parsed text file, from a string literal in a metaprogram, or from a string
computation such as concatenation, slicing, or substitution.
%

\label{sec:structural-strings}
String origins allow us to obtain precise offsets and lengths for
known substrings (e.g., names) so that it is possible to replace
substrings. We use this feature to support \fixCapture for
transformations that produce a target program as a string instead of
an abstract syntax tree. Despite the higher fragility of string-based
transformations, they are common in practice. In our case studies, we use string-based
transformations to generate Java code.

\subsection{Capture-avoiding substitution and inlining}

\begin{figure}[t]
  \begin{subfigure}
    \begin{minipage}[t]{0.45\linewidth}
\begin{lstlisting}[language=simple,style=figureframe]
fun zero() = 0;
fun succ(x) = let n = 1 in x + n;
let n = x + 5 in
  succ(succ(n + x + zero()))
\end{lstlisting}
    \end{minipage}
    \caption{Program with free variable \lstinline[language=simple]!x!.}
  \end{subfigure}
\hfill
\begin{subfigure}
  \begin{minipage}[t]{0.45\linewidth}
\begin{lstlisting}[language=simple,escapeinside=\`\`,style=figureframe]
fun zero() = 0;
fun succ(x) = let n = 1 in (x + n);
let `\highlight{n0}` = 2*n + 5 in 
  succ(succ(`\highlight{n0}` + 2*n + zero()))
\end{lstlisting}
  \end{minipage}
  \caption{Result of substituting \lstinline[language=simple]{2*n} for \lstinline[language=simple]{x}.}
\end{subfigure}
\caption{\fixCapture yields a capture-avoiding substitution that renames
  local variables.}
\label{FIG:captureAvoidingSubst}
\end{figure}

Substitution and inlining are program transformations that may
introduce variable capture. Using \fixCapture, the definition of
capture-avoiding versions of these transformations becomes
straight-forward because \fixCapture takes over the responsibility for
avoiding variable capture. %
Figure~\ref{FIG:captureAvoidingSubst} illustrates the application of
capture-avoiding substitution to a program of a simple language with
global first-order functions and local \emph{let}-bound
variables. 
In the example, we use substitution to replace free occurrences of variable
\lstinline[language=simple]!x! by \lstinline!2*n!. To prevent
capture, our capture-avoiding substitution function renames the locally bound variable~\lstinline!n!.

Substitution is a program transformation where the source
and the target language coincide. Capture-avoiding substitution must
retain the binding structure of the original (source) program. Since
this requirement is part of our definition of capture-avoiding
transformations, we can use \fixCapture to get a capture-avoiding
substitution function from a capturing substitution function. This
simplifies the definition of substitution for our procedural language
to the following:

\begin{lstlisting}[language=rascal,escapeinside=\`\`,style=inlineframe]
subst(p, x, e) = $\fixCapture$(resolve(p), substP(p, x, e));
substP(p, x, e) = prog([substF(f, x, e) | f <- p.fdefs], [substE(e2, x, e) | e2 <- p.main]);
substF(fdef(f, params, b), x, e) = fdef(f, params, x in params ? b : substE(b, x, e));

substE(var(y), x, e)         $\hskip-.1em$ = x == y ? e : var(y);
substE(let(y, e1, e2), x, e) = let(y,  substE(e1, x, e),  x == y ? e2 : substE(e2, x, e));
substE(e1, x, e)             $\hskip.12em$ = for (Exp e2 <- e1) insert substE(e2, x, e);
\end{lstlisting}
%
Function \lstinline!substP! takes a program \lstinline!p! and substitutes
\lstinline!e! for \lstinline!x! in all function definitions and expressions of
the main routine using \lstinline!substF! and \lstinline!substE!,
respectively. Function \lstinline!substF! substitutes \lstinline!e! for
\lstinline!x! in the body of a function only if \lstinline!x! does not occur as
parameter name of the function, that is, only if \lstinline!x! is indeed free in
the function body. Function \lstinline!substE! proceeds similarly for
\emph{let}-bound variables. The final case of \lstinline!substE! uses Rascal's
generic-programming features~\cite{KlintSV09} to provide a default
implementation: We substitute \lstinline!e! for \lstinline!x! in each direct
subexpression of \lstinline!e1! and insert the corresponding result in place of
the subexpression.

Function \lstinline!subst! ensures capture avoidance, but function
\lstinline!substP! does not: When pushing expression \lstinline!e!
under a binder, the bound variable may occur free in \lstinline!e!, in
which case the bound variable should be renamed. By using \fixCapture,
we can omit checking and potentially renaming the bound variable both
for function definitions and for \emph{let} expressions and still get
a capture-avoiding substitution function \lstinline!subst! that
behaves as illustrated in Figure~\ref{FIG:captureAvoidingSubst}.

\begin{figure}[tp]
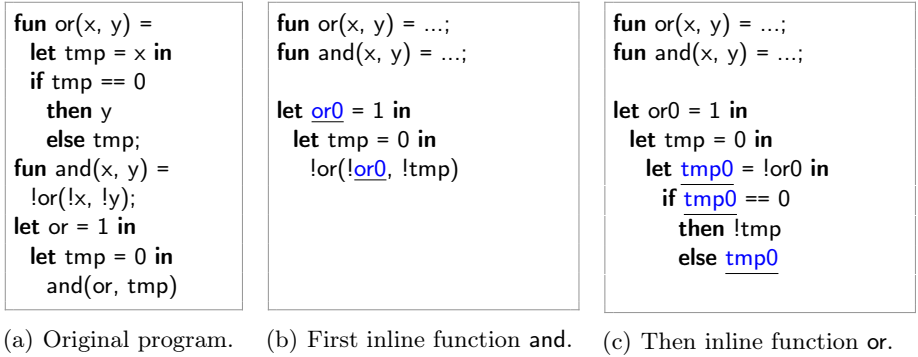

  \begin{subfigure}
    \begin{minipage}[t]{0.24\linewidth}
\begin{lstlisting}[language=simple,style=figureframe]
fun or(x, y) =
  let tmp = x in 
  if tmp == 0 
    then y
    else tmp;
fun and(x, y) = 
  !or(!x, !y);
let or = 1 in
  let tmp = 0 in
    and(or, tmp)
\end{lstlisting}
    \end{minipage}
    \caption{Original program.}
  \end{subfigure}
\hfill
\begin{subfigure}
  \begin{minipage}[t]{0.32\linewidth}
\begin{lstlisting}[language=simple,escapeinside=\`\`,style=figureframe]
fun or(x, y) = ...; 
fun and(x, y) = ...;

let `\highlight{or0}` = 1 in
  let tmp = 0 in
    !or(!`\highlight{or0}`, !tmp)



˚
\end{lstlisting}
  \end{minipage}
  \caption{First inline function \lstinline!and!.}
\end{subfigure}
\hfill
\begin{subfigure}
  \begin{minipage}[t]{0.32\linewidth}
\begin{lstlisting}[language=simple,escapeinside=\`\`,style=figureframe]
fun or(x, y) = ...; 
fun and(x, y) = ...;

let or0 = 1 in
  let tmp = 0 in
    let `\highlight{tmp0}` = !or0 in
      if `\highlight{tmp0}` == 0
        then !tmp
        else `\highlight{tmp0}`
˚
\end{lstlisting}
  \end{minipage}
  \caption{Then inline function \lstinline!or!.}
\end{subfigure}
\caption{Capture-avoiding function inlining is similar to hygienic macro expansion.}
\label{fig:inling-example}
\end{figure}

\def\isimple#1{\lstinline[language=simple]{#1}}

Inlining of functions is a common program-optimization technique used by
compilers. We illustrate our implementation of capture-avoiding inlining in
Figure~\ref{fig:inling-example}. The left column shows a simple program
using two logical functions \isimple{or} and \isimple{and}. The central column
shows the program after inlining \isimple{and}. Note that our language uses a
single namespace for functions and \emph{let}-bound variables. We avoid capture of the
reference to \isimple{or} by renaming the local variable
\isimple{or} to \isimple{or0}. The right column shows the result of inlining
\isimple{or} in the central program. The local variable \isimple{tmp} in the definition of \isimple{or} is
renamed to \isimple{tmp0} since otherwise it would capture the reference to the variable
\isimple{tmp} of the main body.

Based on our implementation of substitution, we can easily implement inlining by
calling \lstinline!substE! to substitute all arguments of a function call into
the body of the function. Like for substitution, it suffices to call \fixCapture
\emph{after} function inlining is complete. Intuitively, this is because
\fixCapture only renames bound variables, which are ignored by
\lstinline!substE! anyway. A detailed investigation of \emph{when} to call \fixCapture 
is part of our future work.

\subsection{Lambda lifting}

Language extensions augment a base language with additional language
features. Many compilers first \emph{desugar} a source program to a
core language. Extensible languages like
SugarJ~\cite{Erdweg13thesis} enable regular programmers to
define their own extensions via custom desugaring
transformations. Such desugaring transformations should preserve the
binding structure of the source program. In fact, the lack of
capture-avoiding and hygienic transformations in extensible languages
was a major motivation of this work.

Exemplary, to show that \fixCapture supports language extensions, we
implemented an extension of our procedural language for local function
definitions that we desugar by lifting them into the
global toplevel function scope~\cite{Johnsson85}. The left column of
Figure~\ref{FIG:lambdaLift} shows an example usage of the extension,
where we have a global function \lstinline!f! that is shadowed by a
local function \lstinline!f!, which is used in another local function
\lstinline!g!. When lifting the two local functions, we get two
toplevel functions named \lstinline!f!, where the originally local
\lstinline!f! captures a call to the originally global \lstinline!f! in the
definition of \lstinline!y!.
Accordingly, \fixCapture renames the lifted function \lstinline!f! and
its calls, both in the main program and the lifted version of
\lstinline!g!.

We implement lambda lifting by recursively (i) finding local functions, (ii)
adapting calls to the local function to pass along variables that occur free in
the function body, and (iii) lifting the function definition to the toplevel. To
identify calls of a local function, we use the name graph of the non-lifted
program. A single call to \fixCapture after desugaring suffices to eliminate
potential name shadowing between functions in the toplevel function scope.




\begin{figure}[tp]
  \begin{subfigure}
    \begin{minipage}[t]{0.45\linewidth}
\begin{lstlisting}[language=simple,style=figureframe]
fun f(x) = x + 1;
let y = f(10) in
  let fun f(x) = f(x + y) in
    let fun g(x) = f(y + x + 1) in
      f(1) + g(3)
\end{lstlisting}
    \end{minipage}
    \caption{Example with local functions \lstinline!f! and \lstinline!g!.}
  \end{subfigure}
\hfill
%
\begin{subfigure}
  \begin{minipage}[t]{0.45\linewidth}
\begin{lstlisting}[language=simple,escapeinside=\`\`,style=figureframe]
fun f(x) = x + 1;
fun `\highlight{f0}`(x, y) = `\highlight{f0}`(x + y, y);
fun g(x, y) = `\highlight{f0}`(y + x + 1, y);
let y = f(10) in
  `\highlight{f0}`(1, y) + g(3, y)
\end{lstlisting}
  \end{minipage}
  \caption{Desugaring of local functions.}
\end{subfigure}
\caption{Lambda lifting of local functions \lstinline!f! and
  \lstinline!g! requires renaming to avoid capture.}
\label{FIG:lambdaLift}
\end{figure}

\subsection{State machines}

\begin{figure}[tp]
\begin{lstlisting}[language=rascal,style=figureframe,escapeinside=\`\`,deletekeywords={map}]
list[FDef] compile(list[State] states) =
  map(state2const, states) + map(state2dispatch, states) + mainDispatch(states)

FDef state2const(State s, int i) =
  fdef(`\highlight{s.name}`, [], val(nat(i)));
FDef state2dispatch(State s) =
  fdef(`\shighlight{"<s.name>-dispatch"}`, ["event"], transitions2cond(s.transitions, val(error())));
Exp transitions2cond([t, *ts], Exp els) =
  cond(equ(var("event"), val(string(t.event)))
        , call(`\highlight{t.state}`, [])
        , transitions2cond(ts, els));
FDef mainDispatch(states) = 
  fdef("main", ["state","event"], mainCond(states, val(error())))
Exp mainCond([s, *ss], Exp els) =
  cond(equ(var("state"), call(`\highlight{s.name}`, []))
        , call(`\shighlight{"<s.name>-dispatch"}`, [var("event")])
        , mainCond (ss, els));
\end{lstlisting}
\caption{Implementation of compiler from state machines to our procedural language.}
\label{FIG:stm2simple}
\end{figure}

In Section~\ref{sec:introduction}, we introduced a language for
state machines to illustrate the problem of inadvertent capture in
program transformation. The \fixCapture algorithm can be used to
repair the result of the transformation without changing the
transformation itself. As a result, developers can structure
transformations in almost arbitrary ways. In the case of the
state-machine compiler, a simple naming convention suffices to link
generated references to declarations. In our case study, the
conventions are that state names become constants and state names
suffixed with \lstinline!-dispatch! become dispatch functions.

We believe the increased liberty of using naming conventions
simplifies the implementation of program transformations. We
illustrate the main part of the compiler of state machines to our
procedural language in Figure~\ref{FIG:stm2simple}. In contrast to approaches based on
explicit binders such as HOAS~\cite{PfenningE88} or
FreshML~\cite{ShinwellPG03}, generated references do not have to
literally occur below their binders in the transformation itself. For
example, function \irascal{compile} independently generates state
constants, state dispatch functions, and the main dispatch function
(by \lstinline!mainCond!), even though the main dispatch function refers to both
generated constants and state dispatch functions via naming
conventions.

\begin{figure}[tp]
  \begin{subfigure}
    \begin{minipage}[t]{0.25\linewidth}
\begin{lstlisting}[language=statemachine,style=figureframe]
state current
  close => closed 
end
  
state closed
  open => current
  lock => token 
end 

state token
  unlock => closed 
end
˚
\end{lstlisting}
    \end{minipage}
    \caption{Renamed door state machine.}
  \end{subfigure}
\hfill
\begin{subfigure}
  \begin{minipage}[t]{0.7\linewidth}
\begin{lstlisting}[language=java,escapeinside=\`\`,style=figureframe]
public class Door {
  final int `\highlight{current}` = 0, closed = 1, token = 2;  
  void run(...) {
    int current0 = `\highlight{current}`;  String token0 = null;
    while ((token0 = input.nextLine()) != null) {
      if (current0 == `\highlight{current}`)
       {if (close(token0)) current0 = closed; else continue;}
      if (current0 == closed)
       {if (open(token0)) current0 = `\highlight{current}`;
        else if (lock(token0)) current0 = token; else continue;}
      if (current0 == token)
       {if (unlock(token0)) current0 = closed; else continue;}
}}}
\end{lstlisting}
  \end{minipage}
  \caption{Renaming of local variables \lstinline!current! and \lstinline!token! to preserve the references of the state machine (exemplarily highlighted).}
\end{subfigure}
\caption{Application of \fixCapture for generated Java code with JDT name resolution.}
\label{FIG:statemachineJava}
\end{figure}

\paragraph{Compilation to Java.}
To exercise capture-avoiding transformation in a more realistic
setting, we also applied \fixCapture on the result of compiling state
machines to Java. To obtain a name graph for Java, we used Rascal's
M$^3$ source code model of Java, which provides accurate name and type
information extracted from the Eclipse JDT~\cite{IzmaylovaKSV13}. The
compiler from state machines to Java generates Java code as structural
strings (cf.~Section~\ref{sec:structural-strings}). It generates a
constant for each state and a single dispatch loop in a \ijava{run}
method.

We illustrate the application of the compiler and the use of \fixCapture on the
generated Java code in Figure~\ref{FIG:statemachineJava}. The left
column shows the state machine from
Figure~\ref{fig:door-state-machine} where we consistently renamed
states \lstinline!opened! and \lstinline!locked! to
\lstinline!current! and \lstinline!token!, respectively. The right
column shows the compiled Java program. Since the dispatch loop in
\ijava{run} uses \lstinline[language=java]{current} to store the
current state and \lstinline[language=java]{token} to save the
last-read token, the compilation introduces variable capture. Note that even without using \fixCapture, the
generated code compiles fine but is ill-behaved because
\lstinline!current==current! in the first \emph{if} would always
succeed. \fixCapture repairs the variable capture by renaming the
local variables.  This case study shows that \fixCapture and our
implementation are not limited to simple languages, but are
applicable for generating capture-free programs of 
languages like Java.

\subsection{Digital forensics with \Derric}

\Derric is a domain-specific language for describing (binary) file
formats~\cite{BosS11}. Such descriptions are used in digital forensic
investigations to recover evidence from (possibly damaged) storage
devices. \Derric descriptions consist of two parts. The first part
describes the high-level structure of a file format by listing
sequence constraints on basic building blocks (called structures) of a
file. The second part describes each structure by declaring fields,
their type, and inter-structure data dependencies. From these
descriptions, the \Derric compiler generates high-performance
validators in Java that check whether a byte sequence matches the
declared format.

\begin{figure}[tp]
  \begin{subfigure}
    \begin{minipage}{0.25\linewidth}
\begin{lstlisting}[language=derric,style=figureframe,literate={{˚}{\texttt{\ }}{1}}]
format Bad

sequence S1 S2

structures
S1 { x: 0x0;  y: S2.x; }
S2 { x; }





˚
\end{lstlisting}
    \end{minipage}
    \caption{A \Derric format.}
  \end{subfigure}
\hfill
%
\begin{subfigure}
  \begin{minipage}{0.7\linewidth}
\begin{lstlisting}[language=java,escapeinside=\`\`,style=figureframe]
public class Bad {
  private long x;
  private boolean S1() {
    markStart();
    long `\highlight{x0}` = ...;  ValueSet vs2 = ...;
    vs2.addEquals(0);
    if (!vs2.equals(`\highlight{x0}`)) return noMatch();
    long y = ...;  ValueSet vs5 = ...;
    vs5.addEquals(x); 
    if (!vs5.equals(y)) return noMatch();
    addSubSequence("S1");
    return true;
}...}
\end{lstlisting}
  \end{minipage}
  \caption{The local variable shadows the field and must be renamed.}
\end{subfigure}
\caption{\fixCapture eliminates variable capture for existing DSL
  compiler of \Derric.}
\label{FIG:derric}
\end{figure}

We show a minimalist, artificial \Derric format description in the
left column of
Figure~\ref{FIG:derric}. The format declares two structures
(\lstinline[language=derric]{S1} and \lstinline[language=derric]{S2}),
which must occur in sequence. \lstinline[language=derric]{S1}
contains two fields: \lstinline[language=derric]{x}, which must be 0,
and \lstinline[language=derric]{y}, which should be equal to field
\lstinline[language=derric]{x} of \lstinline[language=derric]{S2}, which is not
further constrained.
We show an excerpt of the code generated by the \Derric compiler in the right
column of Figure~\ref{FIG:derric}. The main issue is in method
\lstinline[language=java]{S1}, which handles format recognition of structure
\lstinline[language=derric]{S1}. Field~\lstinline[language=java]{x}, which \Derric uses to communicate
\lstinline[language=derric]{S2}'s field~\lstinline[language=derric]{x} to method \lstinline[language=java]{S1}
is shadowed by the local variable~\lstinline[language=java]{x} which
corresponds to \lstinline[language=derric]{S1}'s
field~\lstinline[language=derric]{x}. Without going into too much detail, it is instructive to note that
the Java code compiles fine even without any renaming, but it behaves
incorrectly: Instead of checking \lstinline!S1.y = S2.x!, it checks
\lstinline!S1.y = S1.x!. Such scenario occurs whenever two
structures have a field of the same name and one structure access this field of
the other structure in a constraint. \fixCapture restores correctness by consistently renaming
the local variable in case of capture.

The \Derric case study illustrates the flexibility and power of
\fixCapture. \Derric is a real-world DSL compiling to a mainstream
programming language (Java). The compiler consists of multiple
transformations for desugaring and optimization. The result of these
transformations is an intermediate model of a validator, which is then
pretty printed to Java. Nevertheless, we did not have to modify the
\Derric compiler in any significant way to be able to repair
inadvertent captures, nor was the compiler designed with \fixCapture
in mind. This is shows that our approach is readily
applicable in realistic settings.

\section{Discussion}
\label{SECT:discussion}

We reflect on the problem statement of this work, explain how
\fixCapture supports breaking hygiene, and point out open issues and
future work.


\paragraph{Problem statement.}
In section~\ref{SECT:problemstatement}, we postulated five design
goals for \fixCapture, all of which it satisfies. In
Section~\ref{sec:formalization}, we have verified that \fixCapture
preserves reference intent~(G1) and declaration extent~(G2) of the
source program. Moreover, we have established an equivalence theory for
\fixCapture that at least supports noninvasiveness~(G3). In the
previous section, we have shown how \fixCapture can be applied in a
wide range of scenarios using different languages: state machines,
a simple procedural language, \Derric, and Java.  These results support our
claim that capture elimination with \fixCapture is language-parametric~(G4).

Although the case studies are all implemented in Rascal, any
transformation engine that propagates the unique labels of names is
suited for \fixCapture. Similar to our encoding, one could easily
imagine representing names as tagged strings
\lstinline!Name = (String,Int)!. A structural representation of
strings or compound identifiers are not necessary. Moreover, we do not
require that transformations are written in any specific style to
support capture elimination. In particular, our transformations make
use of sophisticated language features such as intermediate open terms
or generic programming. We conclude that a mechanism like \fixCapture
is transformation-parametric and realizable in other transformation
engines~(G5).


\paragraph{Breaking hygiene.}
Some transformations require that source programs refer to names
synthesized by a transformation. Such breaking of hygiene often occurs
with implicitly declared variables. In other words, intended capture
implies that there is a source reference that is not bound by a
declaration in the source program. Consider, \emph{anaphoric
conditionals} which are like normal \emph{if}-expressions but allow
reference to the result of the condition using a special variable
\isimple{it}~\cite{BarzilayCF11}. For instance, in the expression
\isimple{aif c then !it else it}, the variable \isimple{it} implicitly
refers to a local variable generated by the desugaring of
\isimple{aif}. Applying \fixCapture, however, resolves the capture
which in this case is intended: \isimple{let it0 = c in if it0 then
  !it else it}. To break hygiene in such cases, the transformation
must mark the source occurrences of \isimple{it} when they are carried
over to the result: 
\lstinline[language=rascal,escapeinside=\`\`]{aif(c, t, e) => let("it", c, cond(var("it"), $mark$("it", t), $mark$("it", e)))}.
In our implementation, $mark(s, t)$ sets a
\irascal{synthesized=true} attribute on the \irascal{ID} of any string
$s$ in $t$. Effectively this means that such names are treated as
synthesized names instead of source names.  As a result, \fixCapture
does not rename the binder, and the result of desugaring the above expression
will be \isimple{let it = c in if it then !it else it}.

\paragraph{Future work.}
Theorem~\ref{thm:name-fix-hygiene} shows that \fixCapture turns
sub-hygienic transformations into hygienic transformations. However,
there is currently no decision procedure for whether a transformation
is sub-hygienic or not. For a Turing-complete metalanguage, a static
analysis can only approximate this property. Nevertheless, a
conservative analysis would be useful as it can \emph{guarantee} that
a transformation is sub-hygienic. For example, all transformations of
our case studies except substitution are sub-hygienic, but we have not
formally ensured that. We expect a type system that checks sub-hygiene
to provide guidance to transformation developers similar to
FreshML~\cite{ShinwellPG03}, but without reducing the flexibility.

Another open issue is \textit{when} to apply \fixCapture. This is important when
building transformations on top of other transformations or composing
transformations sequentially into transformation pipelines. After every
application of a transformation, there could be inadvertent variable capture
that \fixCapture can eliminate. For our case studies we used informal reasoning
to decide whether the call to \fixCapture can be delayed, but more principled
guidance would be useful. For example, a simple class of transformations that
commutes with applications of name-fix is the class of name-insensitive
transformations, such as constant propagation.  More generally, care has to be
taken whenever a transformation compares two names for equality, because
intermediate variable capture may yield inaccurate equalities. Since name-fix is
the identity on capture-free programs (Theorem~\ref{thm:fixCapture-base-id}), applying name-fix more than
necessary is at most inefficient, but not unsafe. 


\fixCapture renames not only synthesized names but also names that
originate from the source program. This may break the expected
interface of the generated code. Accordingly, \fixCapture currently is
a whole-program transformation that does not support linking of
generated programs against previously generated libraries, because
names in these libraries cannot be changed. Therefore, \fixCapture is
currently ill-suited for separate compilation. We have experienced this problem in the \Derric compiler, where a \Derric
field named \irascal{BIG_ENDIAN} will shadow a constant with the same
name that occurs in \Derric's precompiled run-time system. We leave the
investigation of a modular \fixCapture for future work.

 
Finally, the current implementation of \fixCapture requires repeated
execution of the name analysis of the target language. As a result,
\fixCapture can be expensive in terms of run-time performance. When a
compiler is run continuously in an IDE, this penalty can be an
impediment to usability. Fortunately, incremental name analysis is a
well-studied topic
(e.g.,~\cite{RepsTD83,WachsmuthKVGV13})
that is likely to yield benefits for \fixCapture because (i)~we know
the delta induced by \fixCapture (renamed variables) and (ii)~new
variable capture can only occur in references that have changed.


\section{Related work}

Various approaches to ensuring capture avoidance have been studied in previous
work.
%
%
Many of them
represent a program not as a syntax tree, but use the syntax tree as a
spanning tree for a graph-based program representation with
additional links from variable references to the corresponding
variable declarations. The advantage of graph-based representations is
that variable references are unambiguously resolved at all times,
which can guide developers of transformations. For example, nameless
program representations such as de~Bruijn indices~\cite{deBruijn72}
encode the graph structure of variable bindings via numeric values;
Oliveira and L\"oh directly encode recursion and sharing in the
abstract syntax of embedded DSLs~\cite{OliveiraL13} via
structured graphs. The disadvantage of these techniques is that they
require explicit handling of graphs (updating indices, redirecting
edges) and do not support open terms well.

In higher-order abstract
syntax~(HOAS)~\cite{PfenningE88} variable references
and declarations are encoded using the binding constructs of the
metalanguage. Thus, developers of transformations inherit name
analysis and capture-avoiding substitution from the metalanguage and
work with fully name-resolved terms. It is well-known that HOAS has a
number of practical problems~\cite{Sheard01}. For
instance, the use of metalevel functions to encode binders makes them
opaque; it is not possible to represent open terms or to pattern match
against variable binders inside constructs such as \lstinline!let!.

FreshML~\cite{ShinwellPG03} uses types to describe the binding structure of
object-language variable binders. This enables deconstruction of a
variable binder via pattern matching, which yields a fresh name and
the body as an open term in which the bound variable has been renamed
to the fresh one. Due to using fresh variables, accidental variable
capture cannot occur but intentional variable capture is possible.
FreshML is limited by using types for declaring variable scope,
because this is only possible for ``declare-before-use''
lexical scoping and not, for example, for the scoping of methods
in an object-oriented class.



In model-driven engineering it is common to describe
abstract syntax using class-based metamodels~\cite{PaigeKP12}.
Syntactic categories correspond to classes, parent-child relations and
cross-references are encoded using associations. Metamodels are expressive
enough to model programs with each name resolved to its declaration
using direct references (pointers). As a result, a large class of
model-transformation formalisms are based on graph
rewriting~\cite{CzarneckiH06}. However, we are unaware of any
work in this area that addresses capture
avoidance. Especially, in the case of model-to-text (M2T)
transformations, names have to be output and all guarantees about
capture avoidance (if any) are lost.

Seminal work on hygiene has been performed in the context of syntax
macros~\cite{KohlbeckerFFD86,ClingerR91}. Like \fixCapture, hygienic macro
expansion automatically renames bound variables to avoid variable capture.  In
related work, a number of approaches to hygienic macro expansion have been
proposed~\cite{Bawden1988,ClingerR91,DybvigHB92,Herman12}. Closest to our work
is the expansion algorithm proposed by Dybvig, Hieb, and
Bruggeman~\cite{DybvigHB92} in that they also associate additional contextual
information to identifiers in syntax objects, similar to our string
origins. However, in their work renamings appear during macro expansion (modulo
lazy evaluation), whereas we perform renamings after transformation. Moreover,
since for macros the role of an identifier only becomes apparent after macro
expansion, they have to track alternative interpretations for a single
identifier. In contrast, we require name analysis for the source language, which
enables a completely different approach to hygienic transformations.

Marco~\cite{LeeGHM12} is a language-agnostic macro engine that detects variable
capture by parsing error messages produced by an off-the-shelve compiler of the
base language. Marco checks whether any of the free names introduced by a macro
is captured at a call-site of the macro. While Marco does not require name
analysis, it has to rely on the quality of error messages of the base compiler,
provides no safety guarantees, and can only detect but not fix variable capture.

Generation environments~\cite{SmaragdakisB99} are metalanguage values
that allow the scoping of variable names generated by a program
transformation. A program transformation can open a generation
environment to generate code relative to the encapsulated lexical
context. Since generation environments can be passed around as
metalanguage values, different transformations can produce code for a
shared a lexical context. While generation environments simplify the
implementation of transformations, they rely on the discipline of
developers and do not provide static guarantees.


Another area where capture avoidance is important is rename
refactorings. In particular, previous work on rename refactoring for
Java~\cite{SchaferEM08} omits checking preconditions and
instead tries to fix the result of a renaming through qualified names
so that reference intent is preserved. De~Jonge~et~al. generalize this
approach to support name-binding preservation in refactorings for
other languages~\cite{JongeV12}. In contrast to our work,
rename refactorings are a limited class of transformations
that do not introduce any synthesized names.

\section{Conclusion}

We presented \fixCapture, a generic solution for eliminating variable
capture from the result of program transformations by comparing name
graphs of the transformation's input and output. This work brings
benefits of hygienic macros to the domain of program
transformations. In particular, \fixCapture relieves developers of
transformations from manually ensuring capture avoidance, and it
enables the safe usage of simple naming conventions.
We have verified that \fixCapture terminates, is correct, and yields
$\alpha$-equivalent programs for inputs that are equal up to possibly
capturing renaming. As we demonstrated with case studies on program
optimization, language extension, and DSL compilation, \fixCapture is
applicable to a wide range of program transformations and
languages.

\paragraph{\ackname} We thank Mitchel Wand, Paolo Giarrusso, Justin Pombrio, Atze van der Ploeg,
and the anonymous reviewers for helpful feedback.

\bibliographystyle{abbrv}



\bibliography{bib}

\iftechreport

\clearpage
\appendix
\chapter*{Appendix}

\section{Proofs of theorems from Section~\ref{sec:formalization} and
Section~\ref{sec:hygienic-transformations}} \label{app:proofs}

\setcounter{theorem}{0}
\setcounter{lemma}{0}

\lemmaOfTermination
\proofOfTermination

\lemmaOfCaptureAvoidance
\proofOfCaptureAvoidance

\lemmaOfSubAlphaEquiv
\proofOfSubAlphaEquiv

\lemmaOfFixCaptureBaseId
\proofOfFixCaptureBaseId

\lemmaOfRenamingPreservesStructure
\proofOfRenamingPreservesStructure

\lemmaOfFixCaptureBaseAlpha
\proofOfFixCaptureBaseAlpha

\lemmaOfFixCaptureStepPreservesSubAlpha
\proofOfFixCaptureStepPreservesSubAlpha

\lemmaOfFixCapturePreservesSubAlpha
\proofOfFixCapturePreservesSubAlpha

\lemmaOfFixCaptureInnerInduction
\proofOfFixCaptureInnerInduction

\lemmaOfFixCaptureSubAlphaInvariant
\proofOfFixCaptureSubAlphaInvariant

\lemmaOfNameFixHygiene
\proofOfNameFixHygiene

\fi

\end{document}